\documentclass[pra,a4paper,aps,twocolumn,superscriptaddress,longbibliography]{revtex4-2}

\usepackage{amsmath, empheq,braket,amsthm,amssymb,graphics,amsfonts,array,color,subfigure, mathrsfs, dsfont, physics, float}
\definecolor{myurlcolor}{rgb}{0,0,0.7}
\definecolor{myurlcolor1}{rgb}{0,0.7,0.1}
\definecolor{myrefcolor}{rgb}{0,0,0.7}
\usepackage[unicode=true,pdfusetitle, bookmarks=true,bookmarksnumbered=false,bookmarksopen=false, breaklinks=false,pdfborder={0 0 0},backref=false,colorlinks=true, linkcolor=myrefcolor,citecolor=myurlcolor1,urlcolor=myurlcolor]{hyperref}
\usepackage{lipsum,MnSymbol}
\usepackage[utf8]{inputenc}  
\usepackage[T1]{fontenc}    
\usepackage{hhline}
\usepackage{soul}

\newtheorem{definition}{Definition}
\newtheorem*{definition*}{Definition}
\newtheorem{theorem}{Theorem}
\newtheorem{lemma}{Lemma}
\newtheorem*{theorem*}{Theorem}
\newtheorem{conjecture}{Conjecture}
\newtheorem*{conjecture*}{Conjecture}

\begin{document}

%\title{On the Wigner entropy of beam splitter states}
\title{Wigner entropy conjecture and the interference formula in quantum phase space}

\author{Zacharie Van Herstraeten}
%\email{zvh@arizona.edu}
\affiliation{Wyant College of Optical Sciences, The University of Arizona, 1630 E. University Blvd., Tucson, AZ 85721}
\affiliation{DIENS, École Normale Supérieure, PSL  University, CNRS, INRIA (QAT), 45 rue d’Ulm, Paris, 75005, France}

\author{Nicolas J. Cerf}
%\email{ncerf@ulb.ac.be}
\affiliation{Centre for Quantum Information and Communication, \'Ecole polytechnique de Bruxelles, CP 165/59, Universit\'e libre de Bruxelles, 1050 Brussels, Belgium}
\affiliation{Wyant College of Optical Sciences, The University of Arizona, 1630 E. University Blvd., Tucson, AZ 85721}

\begin{abstract}
    Wigner-positive quantum states have the peculiarity to admit a Wigner function that is a genuine probability distribution over phase space. The Shannon differential entropy of the Wigner function of such states -- called Wigner entropy for brevity -- emerges as a fundamental information-theoretic measure in phase space and is subject to a conjectured lower bound, reflecting the uncertainty principle. In this work, we prove that this Wigner entropy conjecture holds true for a broad class of Wigner-positive states known as beam-splitter states, which are obtained by evolving a separable state through a balanced beam splitter and then discarding one mode. Our proof relies on known bounds on the $p$-norms of cross Wigner functions and on the interference formula, which relates the convolution of Wigner functions to the squared modulus of a cross Wigner function. Originally discussed in the context of signal analysis, the interference formula is not commonly used in quantum optics although it unveils a strong symmetry under convolution exhibited by Wigner functions of pure states. We provide here a simple proof of the formula and highlight some of its implications. Finally, we prove an extended conjecture on the Wigner-Rényi entropy of beam-splitter states, albeit in a restricted range for the Rényi parameter $\alpha \geq 1/2$.
\end{abstract}

\maketitle

\section{Introduction}

\textit{Quantum phase space.}
The phase-space representation of quantum mechanics has emerged as a powerful and insightful framework for understanding and analyzing quantum systems.
By providing a bridge between the quantum and classical worlds, this approach improves our understanding of quantum systems and opens new research avenues in quantum science. This representation not only provides a more intuitive description of quantum phenomena but also facilitates the study of the quantum-classical correspondence, which is crucial for the development of quantum technologies in particular when dealing with continuous-variable systems.
In short, quantum phase space has proven to be a privileged locus to identify quantum properties and resources \cite{Albarelli2018-ei, Chabaud2021-tq, Lee1991-io, Lutkenhaus1995-gr, De_Bievre2019-se}.

\textit{Wigner negativity.}
Although there exist multiple representations of a quantum state in phase space, the Wigner function appears as the prominent tool for that purpose \cite{Wigner1932-en}.
It indeed benefits from a collection of useful features that make it stand out from its alternatives (see Sec. \ref{sec:preliminaries}).
It serves as the closest quantum analogue to the classical phase-space distribution, offering a complete description of the quantum state while retaining many properties of classical probabilities.
However, unlike classical probability distributions, Wigner functions can take on negative values, reflecting the intrinsic non-classicality of quantum states.
%This feature is particularly significant in exploring fundamental quantum effects such as quantum entanglement, interference, and the behavior of quantum systems under measurement. 
Generally, the negativity of the Wigner function has been widely recognized as a quantum resource \cite{Kenfack2004-cr, Mari2012-fs, Albarelli2018-ei}.
Hence, the Wigner representation effects a fundamental divide between two classes of quantum states: there are Wigner-negative states, which fail to be described by a probability distribution, and Wigner-positive states \footnote{Strictly speaking, states whose Wigner function does not admit any negativity should be denoted as Wigner-non-negative states (since their Wigner function is $\ge 0$ and may even have zeros), but we prefer denoting them as Wigner-positive states in this paper for brevity.}, which admit a fully classical probabilistic description. The focus of this work will be on the latter. Although they may seemingly appear as irrelevant in a study of quantum states since they are devoid of negativity, Wigner-positive states remain subject to the laws of quantum physics and can exhibit intrinsically quantum features (think of the squeezing of light, for example).

\begin{figure}
    \centering
    \includegraphics[width=\linewidth]{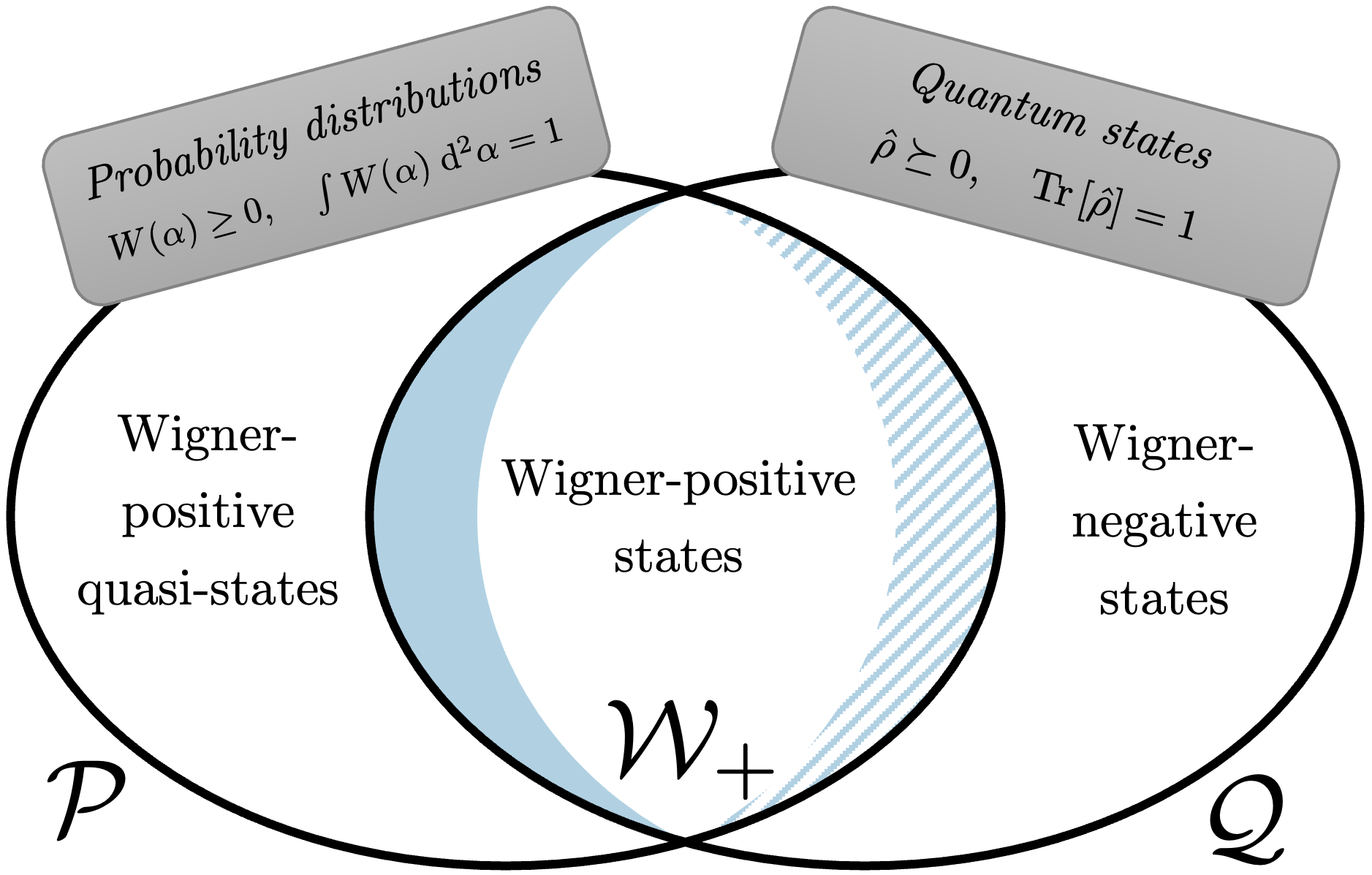}
    \caption{
        Illustrative Venn diagram of the overlap between the set of phase-space probability distributions $W(\alpha)$ and the set of quantum states $\hat{\rho}$.
        The ambient space is the set of Hermitian trace-one operators (not necessarily density operators), corresponding to the set of real normalized functions in phase space (not necessarily probability distributions).
        The set of quantum states $\mathcal{Q}$ contains all positive semi-definite operators $\hat{\rho}$, which are associated with either a Wigner-positive probability distribution or a Wigner-negative \textit{quasi}-probability distribution.
        The set of phase-space probability distributions $W(\alpha)$ is associated to the set of operators $\mathcal{P}$ and contains all Wigner-positive operators, which can be either states or \textit{quasi}-states. The intersection of both sets gives the set of Wigner-positive states $\mathcal{W}_+$. The hatched blue region denotes positive Wigner functions that tend to saturate the positivity condition in phase space (hence becoming \textit{quasi}-probability distributions beyond the border). The solid blue region denotes the set of Wigner-positive states that tend to saturate the positive semi-definiteness operator condition (hence becoming \textit{quasi}-states beyond the border).
    }
    \label{fig:intro_venn}
\end{figure}

\textit{A classical perspective.}
If it was only for classical statistical mechanics, any probability distribution over phase space would define an acceptable state.
In this spirit, we define $\mathcal{P}$ as the set of Hermitian linear operators that are associated with genuine (non-negative) probability distributions in phase space: these are all Wigner-positive trace-one operators.
Note that $\mathcal{P}$ contains positive semi-definite (PSD) operators, and non-PSD operators; we denote the former as states and the latter as quasi-states.
In contrast, the set of quantum states $\mathcal{Q}$ is the set of PSD trace-one operators, which contains Wigner-positive and Wigner-negative operators.
Remarkably, the set of Wigner-positive operators $\mathcal{P}$ and the set of quantum states $\mathcal{Q}$ intersect: their intersection is precisely the set of Wigner-positive quantum states $\mathcal{W}_+=\mathcal{P}\cap\mathcal{Q}$.
As illustrated in Fig.~\ref{fig:intro_venn}, the Wigner-positive set $\mathcal{W}_+$ can be pictured as an intermediate zone at the crossroads of quantum states and classical probability distributions.

\textit{The quantum-classical border.}
In view of this intermediate zone, the border between the sets $\mathcal{Q}$ and $\mathcal{P}$ is two-fold: there is a classical boundary as well as a quantum boundary, as schematically depicted in Fig.~\ref{fig:intro_venn}.
The classical boundary (near the hatched blue region) delimits Wigner-positive and Wigner-negative states; the quantum boundary (near the solid blue region) delimits quantum and non-quantum probability distributions depending on whether the associated trace-one operator is PSD or not.
From a classical point of view, the quantum boundary has nothing special: in fact, it teaches us about the limits of the quantum world. It is linked to the question of when a probability distribution ceases to be a valid Wigner function.
Mathematically, this question has a clear answer: a Wigner function becomes invalid as soon as its corresponding trace-one operator fails to be positive semi-definite \cite{Kastler1965-bv, Loupias1966-vd, Loupias1967-bg, Narcowich1986-kq}.
Yet, this answer can be viewed as unsatisfactory because it lacks a tangible interpretation in phase space. Along this direction, a collection of requirements for physicality have been known:
a valid Wigner function should be continuous and bounded, its covariance matrix is constrained by the uncertainty principe \cite{Robertson1929-js}, and entropic bounds apply to its marginal distributions \cite{Bialynicki-Birula1975-kv, Bialynicki-Birula2007-lp}.
Such necessary conditions are insightful inasmuch as they inform us about the shape of admissible Wigner functions.
Complementing them with any other criteria helps us draw the line with more precision between quantum and non-quantum probability distributions, which is the focus of this work.

\textit{The Wigner entropy.}
Within the set of Wigner-positive operators $\mathcal{P}$, the toolbox of information theory is readily applicable and it makes perfect sense to define the Shannon differential entropy of a phase-space probability distribution, which is nothing else but the Gibbs entropy of statistical mechanics.
Here, we call it the Wigner entropy
\cite{Van_Herstraeten2021-nj}, which captures phase-space uncertainty with an information-theoretic flavor.
From the classical point of view, the Wigner entropy could seemingly be arbitrarily low for highly ``peaked'' states.
However, the situation is different in quantum as it is subject to the uncertainty principle.
Because of that, it is sensible to anticipate that, for quantum Wigner-positive states, the Wigner entropy cannot fall below some critical threshold.
This observation is encapsulated by the Wigner entropy conjecture \cite{Van_Herstraeten2021-nj}, which states a precise lower bound to the Wigner entropy (see Sec. \ref{sec:preliminaries}).
The conjecture has attracted an increasing attention recently \cite{Hertz2017-ta, Van_Herstraeten2021-nj, Van-Herstraeten2023-ww, Cerf2023-gn, Dias2023-qr, Qian2024-vl} but remains open as of today.
As we elaborate in Sec. \ref{sec:preliminaries}, the Wigner entropy is also endowed with other relevant properties, which support its status as physical quantity of its own right.
Provided it is true, the Wigner entropy conjecture unravels a new facet of the quantum-classical border.

\textit{Contribution of this work.}
In this work, we present new evidences towards the validity of the Wigner entropy conjecture. We prove that it holds true for a large family of Wigner-positive states, known as beam-splitter states~\cite{Van_Herstraeten2021-nj} (see Sec.~\ref{sec:beam-splitter-states}).
%We succeed in proving that all beam-splitter states satisfy the Wigner entropy conjecture.
Our result mainly relies on two different ingredients: the interference formula \cite{Janssen1979-es, Hlawatsch1984-yx}, which links the square of a Wigner function to its convolution with itself, and bounds on the $p$-norms of cross Wigner functions \cite{Lieb1990-ev}.
Further, we are able to prove an extended version of the Wigner entropy conjecture formulated in terms of Rényi entropy of the Wigner function (see Sec. \ref{sec:preliminaries}) for beam-splitter states, albeit in some restricted range of the Rényi parameter.

\textit{The interference formula.}
This formula, which is well established in the context of radar ambiguity functions and time-frequency signal analysis \cite{Janssen1979-es, Hlawatsch1984-yx}, seems to be much less studied in a quantum optical context.
Yet, as we shall see, it appears to be a very powerful tool and a fundamental relation obeyed by the Wigner function of any pure state.
It highly constrains the shape of Wigner functions and unveils a ``self-convolution'' symmetry obeyed by all pure states.
In this sense, it is a key ingredient in defining what makes a Wigner function valid.
As it is a central ingredient for our result, which, in our view, warrants broader recognition in the quantum optics community, we devote Sec. \ref{sec:interference_formula} to highlight this formula.

The present paper is organized as follows.
In Sec. \ref{sec:preliminaries} we introduce our notation for the phase-space formalism.
We define the Wigner entropy and present the Wigner entropy conjecture.
We also define the Wigner-Rényi entropy and formulate an extended conjecture.
Sec. \ref{sec:beam-splitter-states} is devoted to beam-splitter states.
In Sec. \ref{sec:interference_formula} we introduce the interference formula.
In Sec. \ref{sec:results}, we present our main results.
Finally, we discuss our results and conclude in Sec. \ref{sec:conclusion}.

\section{Preliminaries}
\label{sec:preliminaries}

Here, we lay out our notations and the technical tools that will be used in the following Sections.
We begin with a short overview of the phase-space formalism in terms of Wigner functions.
We then introduce the Wigner entropy and recall the associated Wigner entropy conjecture. Finally, we present a generalization of this conjecture involving the Wigner-Rényi entropy.

\textit{Quantum operators.} We consider an infinite-dimensional Hilbert space $\mathcal{H}=L^2(\mathbb{R})$.
Pure states are denoted as ket-vectors $\ket{\psi}\in\mathcal{H}$ and mixed states are trace-one Hermitian positive semi-definite density operators $\hat{\rho}\in\mathcal{D}(\mathcal{H})$.
We consider the Fock basis $\lbrace\ket{n}\rbrace_{n\in\mathbb{N}}$ with wave functions $\psi_n(x)=(\sqrt{\pi}2^n n!)^{-1/2}H_n(x)\exp(-x^2/2)$ where $H_n$ is the $n$th Hermite polynomial.
We use the usual mode operator $\hat{a}=(\hat{x}+i\hat{p})/\sqrt{2}$ and photon-number operator $\hat{n}=\hat{a}^\dagger\hat{a}$.
We define the displacement operator $\hat{D}(\alpha)=\mathrm{exp}(\alpha\hat{a}^\dagger-\alpha^\ast\hat{a})$ with $\alpha\in\mathbb{C}$ and the parity operator $\hat{\Pi}=(-1)^{\hat{n}}$.
Finally, we denote the displaced parity operator as $\hat{\Pi}(\alpha)=\hat{D}(\alpha)\hat{\Pi}\hat{D}^\dagger(\alpha)$.

\textit{Weyl transform.}
Phase space can equivalently be identified as the set of couples $(x,p)\in\mathbb{R}^2$ or the set of complex numbers $x+ip=\alpha\in\mathbb{C}$. We stick to the latter convention here, so we use complex numbers $\alpha$ as arguments of phase-space distributions.
%although we may switch from one to the other without ambiguity. We write $x+ip$ and $(x,p)$ interchangeably for the argument of a phase-space distribution. 
A common isomorphism between quantum operators and phase-space distributions is obtained  via the Weyl transform \cite{Royer1977-he}, which maps any linear operator $\hat{A}:\mathcal{H}\rightarrow\mathcal{H}$ to a distribution $A:\mathbb{C}\rightarrow\mathbb{C}$.
The Weyl transform $\hat{A}\mapsto A$ and its inverse transform $A\mapsto\hat{A}$ are defined as \footnote{Note the factor $\sqrt{2}$ in Eqs. \eqref{eq:weyl_transform} and \eqref{eq:inverse_weyl_transform}, which originates from our asymmetric conventions, namely $\hat{a}=(\hat{x}+i\hat{p})/\sqrt{2}$ but $\alpha=x+ip$. It implies that the displacement operator $\hat{D}(\alpha)$ shifts the $(x,p)$ coordinates as $x\mapsto x+ \sqrt{2} \,\mathrm{Re(\alpha)}$ and $p\mapsto p + \sqrt{2} \,\mathrm{Im(\alpha)}$. The reason for this choice is that the definition \eqref{eq-def-Wigner-entropy-phase-space} of the Wigner entropy then coincides with its earlier definition in terms of $(x,p)$ coordinates, see \cite{Van_Herstraeten2021-nj}.}:
\begin{align}
    A(\alpha)
     & =
    \frac{1}{\pi}
    \Tr\left[
        \hat{A}\;
        \hat{\Pi}\left(\alpha/\sqrt{2}\right)
        \right],
    \label{eq:weyl_transform}
    \\[1em]
    \hat{A}
     & =
    2\int
    A(\alpha)
    \;
    \hat{\Pi}\left(\alpha/\sqrt{2}\right)
    \dd^2\alpha.
    \label{eq:inverse_weyl_transform}
\end{align}
The Weyl transform establishes a 1-to-1 correspondence between phase-space distributions and quantum operators.
Importantly, the Weyl transform stands out from other quantum phase-space representations as it is an isometry between quantum operators and phase-space distributions.
Indeed, the inner product between two operators $\hat{A},\hat{B}$ with Weyl transforms $A,B$ is preserved:
\begin{align}
    \Tr[\hat{A}\hat{B}]
     & =
    2\pi
    \int
    A(\alpha)B(\alpha) \, \dd^2\alpha.
\end{align}
The above relation is known as the overlap formula or Moyal's identity \cite{Moyal1949-tl}.

\textit{Wigner function.}
The Weyl transform of a density operator $\hat{\rho}$ is known as the Wigner function of $\hat{\rho}$ and is denoted as $W_{\hat{\rho}}(\alpha)$. Since density operators are Hermitian and trace-one operators, Wigner functions are real-valued functions that are normalized to $1$. The Wigner function is the only phase-space quasidistribution that directly contains the marginal distribution of all quadratures.
Indeed, recalling that $\alpha = x + ip$,
the position distribution is given by $\rho_x(x)\vcentcolon=\bra{x}\hat{\rho}\ket{x}=\int W_{\hat{\rho}}(x,p)\dd p$ while the momentum distribution is given by $\rho_p(p)\vcentcolon=\bra{p}\hat{\rho}\ket{p}=\int W_{\hat{\rho}}(x,p)\dd x$.
Since density operators are positive semi-definite, $\rho_x$ and $\rho_p$ are genuine (non-negative) probability distributions.

\textit{Wigner positivity.}
Although quantum states always have non-negative marginal distributions, their Wigner function can be partly negative.
In fact, Hudson's theorem states that the only Wigner-positive pure states are Gaussian pure states \cite{Hudson1974-ll}.
We say that a state $\hat{\rho}$ is Wigner positive iff its Wigner function is non-negative everywhere, i.e. $W_{\hat{\rho}}(\alpha)\geq 0\ \forall\alpha$ and we note the set of Wigner-positive states as $\mathcal{W}_+$.
A state that is not Wigner positive is called Wigner negative. The picture becomes more intricate when considering mixed states, and, as of today, there is no satisfying description of the set of mixed Wigner-positive states
\cite{Garcia-Bondia1988-tr, Brocker1995-rp, Mandilara2009-kn}. Some known subsets of the set of Wigner-positive states are the convex hull of Gaussian states, passive states \cite{Lenard1978-ru, Bastiaans1983-je},  and beam-splitter states \cite{Van_Herstraeten2021-nj}.
We will give an overview of beam-splitter states in Sec. \ref{sec:beam-splitter-states}.

\textit{Wigner entropy.}
Since Wigner-positive states are described by a genuine probability distribution $W_{\hat{\rho}}$ over phase space, the Shannon differential entropy of $W_{\hat{\rho}}$ is mathematically well defined.
\begin{definition}[Wigner\,entropy\,\cite{Van_Herstraeten2021-nj}]
    The Wigner entropy of a Wigner-positive state $\hat{\rho}\in\mathcal{W}_+$ is defined as
    \begin{align}
        h(W_{\hat{\rho}})
        =
        -\iint W_{\hat{\rho}}(\alpha)\ln W_{\hat{\rho}}(\alpha)\dd^2\alpha.
        \label{eq-def-Wigner-entropy-phase-space}
    \end{align}
\end{definition}

As it inherits all properties from the Shannon entropy, the Wigner entropy is arguably the most natural information-theoretic measure of phase-space uncertainty even though it only has a clear meaning for Wigner-positive states. It has, for example, been used in the context of non-equilibrium thermodynamics as a tool to measure the entropy production rate of quantum stochastic processes~\cite{PhysRevLett.118.220601,PhysRevLett.121.160604}. There, thermal states of coupled harmonic oscillators evolve with a quadratic Hamiltonian, which ensures that the states remain Gaussian at all times, hence Wigner-positive. Interestingly, for Gaussian states, the Wigner entropy has been shown to coincide (up to an additive constant) to the quantum Rényi entropy of order~2~\cite{Adesso2012-ok}. Let us also mention other works exploring the Wigner entropy of electrons in atomic systems \cite{Guevara2003-na} or in harmonic oscillator model systems \cite{Laguna2010-ir, Salazar2023-yf}. To date, the Wigner entropy has no established generalization to Wigner-negative states (note that a complex-valued extension of the Wigner entropy has recently been considered \cite{Cerf2023-gn}).

The Wigner entropy is invariant under symplectic transformations (Gaussian unitaries) and it provides a lower bound to the sum of the marginal entropies of position and momentum, i.e. $h(W_{\hat{\rho}})\leq h(\rho_x)+h(\rho_p)$ \cite{Van_Herstraeten2021-nj}. Since the uncertainty principle forbids a simultaneous determination of position and momentum, it is natural to expect a critical minimum value for the Wigner entropy, which is the rationale for the following conjecture.

% \begin{conjecture}[Wigner entropy conjecture \cite{Van_Herstraeten2021-nj}]
% The Wigner entropy of a Wigner-positive state satisfies the following lower bound:
% \begin{align}
% h(W_{\hat{\rho}})\geq\ln\pi+1
% \qquad\forall\hat{\rho}\in\mathcal{W}_+.
% \label{eq:wig_entropy}
% \end{align}
% \label{conj:wig_entropy}
% \end{conjecture}
%  The Wigner entropy conjecture can be understood as a refinement of uncertainty relation $h(\rho_x)+h(\rho_p)\geq\ln\pi+1$ due to Białynicki-Birula and Mycielski \cite{Bialynicki-Birula1975-kv} in the case of Wigner-positive states, as anticipated in Ref.~\cite{Hertz2017-ta}. Indeed, the subadditivity of the Shannon entropy implies that $h(W_{\hat{\rho}})\leq h(\rho_x)+h(\rho_p)$.

%{\color{red}Cite [17] as first statement of the Wigner entropy conjecture.}
\begin{conjecture}[Wigner entropy conjecture \,\cite{Van_Herstraeten2021-nj}]
    The Wigner entropy of a Wigner-positive state satisfies the following lower bound:
    \begin{align}
        h(W_{\hat{\rho}})\geq\ln\pi+1
        \qquad\forall\hat{\rho}\in\mathcal{W}_+.
        \label{eq:wig_entropy}
    \end{align}
    \label{conj:wig_entropy}
\end{conjecture}
This conjecture was initially expressed in Ref. \cite{Hertz2017-ta}, where it was viewed as a refinement of the uncertainty relation $h(\rho_x)+h(\rho_p)\geq\ln\pi+1$ due to Białynicki-Birula and Mycielski \cite{Bialynicki-Birula1975-kv} in the case of Wigner-positive states.
%Indeed, the subadditivity of the Shannon entropy implies that $h(W_{\hat{\rho}})\leq h(\rho_x)+h(\rho_p)$.
Note that the conjectured lower bound \eqref{eq:wig_entropy} is attained for all pure Gaussian states. First, it is immediate to check that it is saturated for the vacuum state, i.e. the Fock state $\ket{0}$, which is associated with the Wigner function \begin{equation}
    W_0(\alpha)
    =
    \frac{1}{\pi}\exp(-\abs{\alpha}^2) ,
\end{equation}
and gives $h(W_0)=\ln\pi+1$.
Then, the invariance of $h(W_{\hat{\rho}})$ under symplectic transformations directly implies that the bound is reached for all pure Gaussian states. Furthermore, the concavity of the Shannon entropy immediately implies that Conjecture~\ref{conj:wig_entropy} holds true for the convex hull of Gaussian states. The key problem, however, is that $\mathcal{W}_+$ contains many more states than the convex hull of Gaussian states. In Ref. \cite{Van_Herstraeten2021-nj}, it was proven that Conjecture~\ref{conj:wig_entropy} holds true for passive states, which is a distinct subset of $\mathcal{W}_+$.
More recent works have shown that the Wigner entropy is lower bounded by $\ln(2\pi)$ \cite{Dias2023-qr} or even $\ln(2\pi)+ S(\hat \rho)$ \footnote{The lower bound $\ln(2\pi)+ S(\hat \rho)$ depends on the von Neumann entropy $S(\hat \rho)$ of state $\hat \rho$. [T. Haas, private communication, 2024].}, and that it satisfies the conjectured lower bound $\ln\pi+1$ for any Wigner-positive state in the qubit space spanned by the Fock states $\ket{0}$ and $\ket{1}$ \cite{Qian2024-vl}. The present work extends the validity of Conjecture~\ref{conj:wig_entropy} to the much wider subset of beam-splitter states within $\mathcal{W}_+$.

\textit{Wigner-Rényi entropy.}
The Shannon entropy is only a prominent special case of a larger family of entropic functionals called Rényi entropies. Let us first define the $p$-norm of a Wigner function as
\begin{align}
    \Vert W\Vert_p
    =
    \left(
    \int
    \abs{W(\beta)}^p\dd^2\beta
    \right)^{\frac{1}{p}}, \qquad p\in\mathbb{R}_+.
\end{align}
Note that $p$-norms are not only defined for (real) Wigner functions but for any complex-valued distribution.
The $1$-norm of a Wigner function is related to its negative volume, i.e. $\Vert W\Vert_1=1+2\mathcal{N}(W)$ where $\mathcal{N}=\int_{W<0}\abs{W(\alpha)}\dd^2\alpha$.
In the limit $p\rightarrow 0$, the $p$-norm is related to the support of $W$ and one defines $\Vert W\Vert_0\vcentcolon=\nu(\mathrm{supp}(W))$, where $\nu$ stands for the Lebesgue measure. Note that Wigner functions always have an infinite support, so $\Vert W\Vert_0$ diverges \cite{Janssen1998-as}. In the limit $p\rightarrow\infty$, the $p$-norm is related to the maximum absolute value of $W$, and one defines $\Vert W\Vert_\infty\vcentcolon=\max\abs{W}$.
Now that the $p$-norms of a Wigner function have been defined, we are in position to define the Wigner-Rényi entropy.

\begin{definition}[Wigner-Rényi entropy \cite{Van_Herstraeten2021-nj}]
    The Wigner-Rényi entropy of a state $\hat{\rho}$ is defined as:
    \begin{align}
        h_\alpha(W_{\hat{\rho}})
        =
        \frac{\alpha}{1-\alpha}
        \ln\Big(
        \Vert W_{\hat{\rho}}\Vert_\alpha
        \big), \qquad \alpha\in \, (0,\infty].
    \end{align}
    \label{def-Wigner-Renyi}
\end{definition}
Note that $h_1(W_{\hat{\rho}})$ is ill-defined but we consider instead the limit $\alpha\rightarrow 1$ which tends to the Wigner entropy, namely $\lim_{\alpha\rightarrow 1}h_{\alpha}(W)=h(W)$ provided $\Vert W\Vert_1=1$ (which holds for Wigner-positive states); the limit diverges when $\Vert W\Vert_1\neq 1$.
We further define the limiting case $h_\infty(W)\vcentcolon=-\ln(\max\abs{W})$.

\begin{conjecture}[Wigner-Rényi entropy conjecture \cite{Van_Herstraeten2021-nj}]
    The Wigner-Rényi entropy of a Wigner-positive state satisfies the following lower bound:
    \begin{align}
        h_\alpha(W_{\hat{\rho}})
        \geq
        \ln\pi
        +
        \frac{\ln\alpha}{\alpha-1}
        \qquad\forall\hat{\rho}\in\mathcal{W}_+ ,\quad\forall\alpha\in \, (0,\infty].
    \end{align}
    \label{conj:wig_rényi_entropy}
\end{conjecture}

Just as for the Wigner entropy, the vacuum state saturates the bound since we have
\begin{align}
    h_\alpha(W_0)
     & =
    \ln\pi+\frac{\ln\alpha}{\alpha-1} , \quad\forall\alpha\in \, (0,\infty],
\end{align}
and we have $h(W_0)=\ln\pi+1$ in the limit $\alpha\rightarrow 1$. The same is true for all pure Gaussian states because $h_\alpha$ is invariant under symplectic transformations for all $\alpha$.
%In the limit $\alpha\rightarrow 1$ the lower bound goes towards $\ln\pi+1$ so 
From this, it is clear that Conjecture~\ref{conj:wig_entropy} is a special case of Conjecture~\ref{conj:wig_rényi_entropy}. Recently, it was shown that the Wigner-Rényi entropy satisfies this lower bound for all Wigner-positive states in the regime $\alpha\geq 2$ \cite{Dias2023-qr}, which provides a partial proof of Conjecture \ref{conj:wig_rényi_entropy} (but does \textit{not} prove Conjecture~\ref{conj:wig_entropy}).

Lastly, a stronger statement based on continuous majorization theory, which implies Conjecture~\ref{conj:wig_rényi_entropy} (and therefore Conjecture~\ref{conj:wig_entropy}), has also been conjectured.
This so-called Wigner majorization conjecture states that the Wigner function of the vacuum $W_0$ continuously majorizes any non-negative Wigner function \cite{Van-Herstraeten2023-ww}.

%Conj. \ref{conj:wig_rényi_entropy} can be rephrased as $h_{\alpha}(W_{\hat{\rho}})\geq h_{\alpha}(W_0) \forall\alpha$ and $\forall\hat{\rho}\in\mathcal{W}_{+}$.

\section{Beam-splitter states}
\label{sec:beam-splitter-states}

Let us give an overview of the family of Wigner-positive states that we call beam-splitter states \cite{Van_Herstraeten2021-nj}.
Remember that a beam splitter couples two modes $\hat{a}_1$ and $\hat{a}_2$ through the Gaussian unitary operator $\hat{U}_\eta\vcentcolon=\mathrm{exp}(\theta(\hat{a}_1^\dagger\hat{a}_2-\hat{a}_1\hat{a}_2^\dagger))$ where $\eta=\cos^2\theta$ is the transmittance \cite{Weedbrook2012-qu}.
In the Heisenberg picture, it performs the map $\hat{a}_1\mapsto\sqrt{\eta}\hat{a}_1+\sqrt{1-\eta}\hat{a}_2$ and $\hat{a}_2\mapsto\sqrt{\eta}\hat{a}_2-\sqrt{1-\eta}\hat{a}_1$.
In the Schrödinger picture, the action of the beam splitter is appropriately described through a rescaling and a convolution of the Wigner functions.

Let us first define these operations. The rescaling operator $\mathcal{L}_s$ acts on a Wigner function as
\begin{align}
    \mathcal{L}_s\left[W\right](\alpha)
    =
    \frac{1}{s^2}
    \,W\left(\frac{\alpha}{s}\right),
\end{align}
where $s\in\mathbb{R}_{>0}$.
For $s<1$ it performs a contraction of $W$ towards the origin of phase space; for $s>1$ it performs a dilation.
The prefactor ensures that the rescaled function remains normalized. Note that a rescaling may be understood as a change of the constant $\hbar$ (so far assumed to be $1$), that is, the change $\hbar\rightarrow\hbar'$ is described by a rescaling $\mathcal{L}_{s}$ where $s=\sqrt{\hbar'/\hbar}$. Second, the convolution of two Wigner functions is defined as:
\begin{align}
    \big(W_1\ast W_2\big)(\alpha)
    =
    \iint
    W_1(\beta)\,W_2(\alpha-\beta)\dd^2\beta.
\end{align}

\begin{figure}
    \centering
    \includegraphics[width=0.95\linewidth]{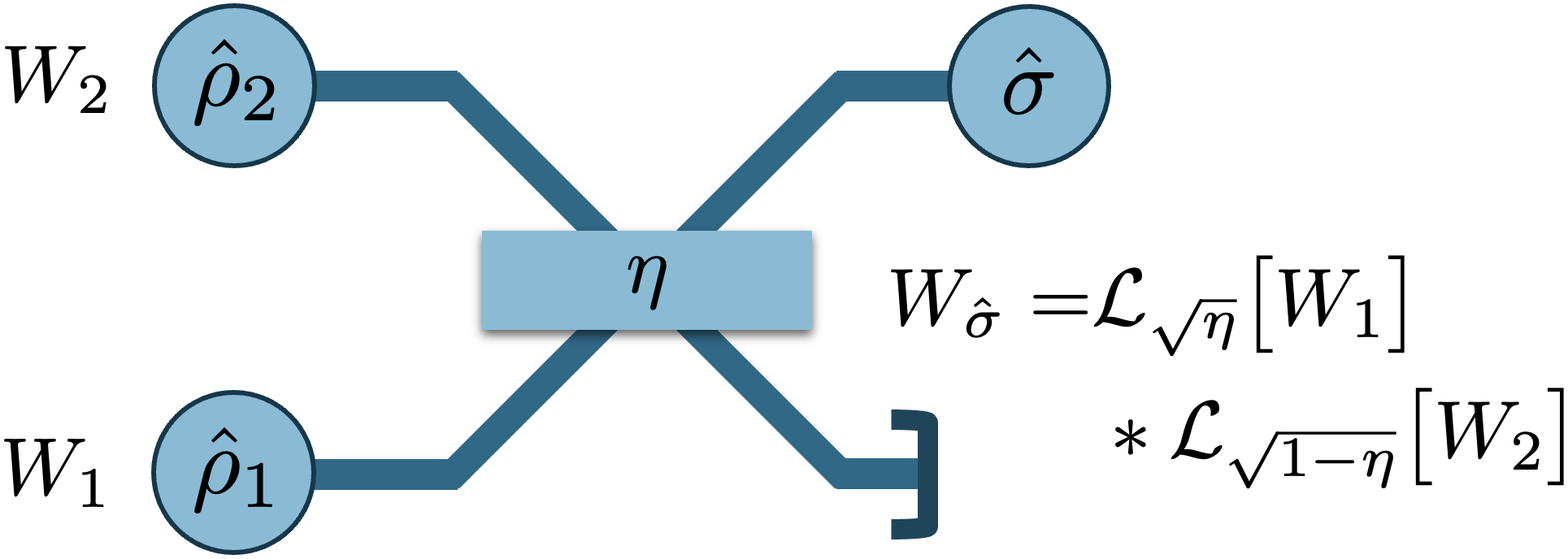}
    \caption{
    Schematic of a beam splitter with transmittance $\eta$ acting on a product input state $\hat{\rho}_1\otimes\hat{\rho}_2$.
    The reduced state of output 1 is $\hat{\sigma}(\hat{\rho}_1,\hat{\rho}_2)=\mathrm{Tr}_2\left[\hat{U}_{\eta}\left(\hat{\rho}_1\otimes\hat{\rho}_2\right)\hat{U}^\dagger_\eta\right]$ and is described by the Wigner function $W_{\hat{\sigma}}=\mathcal{L}_{\sqrt{\eta}}[W_1]\ast\mathcal{L}_{\sqrt{1-\eta}}[W_2]$.
    }
    \label{fig:bs_rho1_rho2}
\end{figure}

With these definitions, the action of the beam splitter can be described as follows. The states $\hat{\rho}_1$ and $\hat{\rho}_2$ (with respective Wigner functions $W_1$ and $W_2$) produce the reduced state $\hat{\sigma}=\Tr_{2}[\hat{U}_\eta(\hat{\rho}_1\otimes\hat{\rho}_2)\hat{U}^\dagger_\eta]$ with Wigner function
\begin{align}
    W_{\hat{\sigma}}
    =
    \mathcal{L}_{\sqrt{\eta}}\big[W_1\big]
    \ast
    \mathcal{L}_{\sqrt{1-\eta}}\big[W_2\big],
    \label{eq:unbalanced_bs}
\end{align}
as illustrated in Fig.~\ref{fig:bs_rho1_rho2}. From the point of view of probability theory, a beam splitter performs the exact equivalent of the scaled addition of two random variables.
More precisely, let $\mathbf{X}=(X_1,X_2)$ be a bivariate random variable described by the probability density distribution $p_{\mathbf{X}}(x_1,x_2)$, and let $\mathbf{Y}$ be similarly defined.
Let $\mathbf{Z}=\sqrt{\eta}\mathbf{X}+\sqrt{1-\eta}\mathbf{Y}$; then we have $p_\mathbf{Z}=\mathcal{L}_{\sqrt{\eta}}[p_\mathbf{X}]\ast\mathcal{L}_{\sqrt{1-\eta}}[p_\mathbf{Y}]$.

The convolution of two Wigner functions $W_1\ast W_2$ is always non-negative \cite{Bertrand1983-ij, Jagannathan1987-yj, Narcowich1988-jr}.
Hence, if we choose the transmittance of the beam splitter to be $\eta=1/2$, Eq.~\eqref{eq:unbalanced_bs} becomes $\smash{W_{\hat{\sigma}}=\mathcal{L}_{1/\sqrt{2}}[W_1\ast W_2]}$ which is a non-negative Wigner function.
This setup producing Wigner-positive states is depicted in Fig.~\ref{fig:bs_rho1_rho2}, setting $\eta=1/2$.
We have restricted to inputs of the product form $\hat{\rho}_1\otimes\hat{\rho}_2$, but observe that any separable input state (i.e., convex mixture of product states) also produces a Wigner-positive state in this setup since a convex mixture of Wigner-positive states is Wigner positive. This leads us to define the set of beam-splitter (BS) states, which we denote as $\mathcal{B}$ \footnote{Note the discrepancy with the notation used in Ref.~\cite{Van_Herstraeten2021-nj}, where $\mathcal{B}$ denotes the subset of states $\hat{\sigma}(\hat{\rho}_1,\hat{\rho}_2)$ whereas $\mathcal{B}_c$ denotes the closure of its convex hull.}.

\begin{definition}[Beam-splitter state]
    A BS state is any state $\hat{\sigma}$ that can be obtained as the single-mode output of a balanced beam splitter acting on a separable input $\hat{\rho}_{\mathrm{sep}}$, i.e.
    \begin{align}
        \hat{\sigma}
        =
        \Tr_2\left[
        \hat{U}_{1/2}\
        \hat{\rho}_{\mathrm{sep}}\
        \hat{U}_{1/2}^{\dagger}
        \right]
        \label{eq:rho_sep_bs_state}
    \end{align}
    where $\hat{\rho}_{\mathrm{sep}}=\sum_i p_i\ket{\psi_i,\varphi_i}\bra{\psi_i,\varphi_i}$, with $p_i\!\ge\!0$ and $\sum_i p_i\!=\!1$.
    \label{def:bs_states}
\end{definition}
In the following, we also use the shorthand notation
\begin{align}
    \hat{\sigma}(\hat{\rho}_1,\hat{\rho}_2)\vcentcolon=\Tr_2\big[\hat{U}_{1/2}(\hat{\rho}_1\otimes\hat{\rho}_2)\hat{U}^{\dagger}_{1/2}\big].
\end{align}
when refering to the BS states produced by the product states
$\hat{\rho}_1\otimes\hat{\rho}_2$.
Note that in the special case where $\hat{\rho}_1=\hat{\rho}_2=\hat{\rho}_G$, with $\hat{\rho}_G$ being an arbitrary Gaussian state, it is well known that $\hat{\sigma}(\hat{\rho}_G,\hat{\rho}_G)=\hat{\rho}_G$, implying that Gaussian states (and any mixture thereof) are included in the set $\mathcal{B}$ of BS states, hence Gaussian states are obviously Wigner positive.

In general, the set $\mathcal{B}$ strictly includes all mixtures of Gaussian states and all passive states \cite{Van_Herstraeten2021-nj}, but it also contains many states that go beyond such mixtures. For this reason, the proof of Conjecture \ref{conj:wig_entropy} for states $\hat\rho \in \mathcal{B}$ as presented in Sec.~\ref{sec:results} is stronger than all earlier proofs. Yet, the BS states are known to be strictly included in the set of Wigner-positive states, i.e., $\mathcal{B}\subsetneq
    \mathcal{W}_+$ \cite{Van_Herstraeten2021-nj}, hence our proof is not a full proof of Conjecture \ref{conj:wig_entropy}.

\begin{figure}
    \centering
    \includegraphics[width=0.95\linewidth]{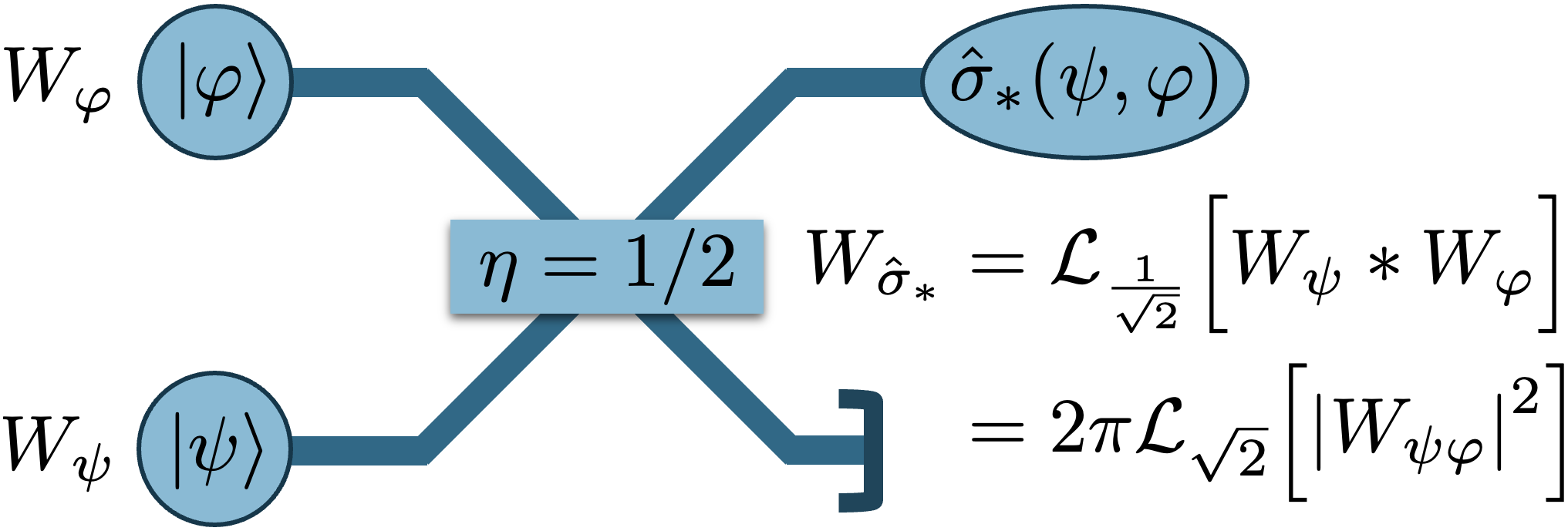}
    \caption{
    Elementary beam-splitter state $\hat{\sigma}_\ast(\psi,\varphi)$ produced from the pure input states $\ket{\psi}$ and $\ket{\varphi}$.
    The output Wigner function is computed as $W_{\hat{\sigma}}=\mathcal{L}_{1/\sqrt{2}}[W_\psi\ast W_\varphi]$, where $W_\psi$ and $W_\varphi$ are the input Wigner functions. As a consequence of the outer interference formula (see Sec.~\ref{sec:interference_formula}), the output Wigner function is also equal to $W_{\hat{\sigma}}=2\pi \, \mathcal{L}_{\sqrt{2}}[\abs{W_{\psi\varphi}}^2]$, where $W_{\psi\varphi}$ is the cross Wigner function.}
    \label{fig:extreme_bs_state}
\end{figure}

\textit{Subset of elementary BS states.} Within the set $\mathcal{B}$, we identify the subset of BS states that are produced by pure separable (product) states, which we call \textit{elementary} BS states and denote with a $\ast$-subscript. They are defined as (see Fig.~\ref{fig:extreme_bs_state})
\begin{align}
    \hat{\sigma}_\ast(\psi,\varphi)
    =
    \Tr_2\left[
        \hat{U}_{1/2}\
        \ket{\psi,\varphi}\bra{\psi,\varphi}
        \hat{U}_{1/2}^{\dagger}
        \right],
    \label{eq:extreme_bs_state}
\end{align}
and their associated Wigner functions are written as
\begin{align}
    W_{\hat{\sigma}_\ast(\psi,\varphi)}=\mathcal{L}_{1/\sqrt{2}}[W_{\psi}\ast W_{\varphi}]  .
    \label{eq:Wigner_fct_extreme_bs_state}
\end{align}
The corresponding subset of elementary BS states is denoted as $\mathcal{B}_{\ast}$.
Observe that any BS state $\hat{\sigma}\in\mathcal{B}$ can be obtained as a convex mixture of BS states $\hat{\sigma}_\ast\in\mathcal{B}_\ast$.
Indeed, just like any separable state admits a (non-unique) decomposition $\hat{\rho}_{\mathrm{sep}}=\sum_{i}p_{i}\ket{\psi_i,\varphi_i}\bra{\psi_i,\varphi_i}$, any BS state admits a (non-unique) decomposition $\hat{\sigma}=\sum_{i}p_{i}\hat{\sigma}_\ast(\psi_i,\varphi_i)$.
By analogy with the set of separable states, the set of BS states $\mathcal{B}$ is the convex hull of $\mathcal{B}_{\ast}$. Due to the concavity of the Shannon entropy, it is therefore sufficient to prove Conjecture~\ref{conj:wig_entropy} for the elements of $\mathcal{B}{\ast}$ in order to establish its validity for all of~$\mathcal{B}$ (as we shall see in Sec.~\ref{sec:results}, the case of Conjecture~\ref{conj:wig_rényi_entropy} is a bit more complicated).
The problem of characterizing the extreme points of $\mathcal{B}$ (which belong to $\mathcal{B}_{\ast}$ but do not exhaust it) is discussed in Appendix~\ref{app-extreme-BS}.

\textit{Subset of Husimi BS states.}
%Subset of $\mathcal{B}$ coinciding with Husimi functions
Within the set $\mathcal{B}$, we may also consider the subset of BS states that are obtained if one input is in the vacuum state. Consider BS states of the form $\hat{\sigma}(\hat{\rho},0)=\Tr_2[\hat{U}_{1/2}(\hat{\rho}\otimes\ket{0}\bra{0})\hat{U}^{\dagger}_{1/2}]$; such states are associated to a Wigner function that is precisely the Husimi function of $\hat{\rho}$, i.e., $W_{\hat{\sigma}}=Q_{\hat{\rho}}$ \cite{Van_Herstraeten2021-nj}, hence we refer to $\hat{\sigma}(\hat{\rho},0)$ as Husimi BS states.
It was proven by Lieb that the entropy of the Husimi function (called the Wehrl entropy) is lower bounded by $\ln\pi+1$ \cite{Lieb1978-gy}, a result which was later generalized by Lieb and Solovej for Rényi entropies \cite{Lieb2014-qx}.
From this, it follows directly that all Husimi BS states satisfy the Wigner entropy conjecture (Conj.~\ref{conj:wig_entropy}) and its generalization (Conj.~\ref{conj:wig_rényi_entropy}).

%Note that if $\ket{\gamma}$ is a Gaussian pure state, $\hat{\sigma}_\ast(\gamma,\gamma)=\ket{\gamma}\bra{\gamma}$, confirming that Gaussian states (and any mixture thereof) are included in the set $\mathcal{B}$.

\section{Interference formula}
\label{sec:interference_formula}

\begin{figure*}
    \centering
    \includegraphics[width=0.95\linewidth]{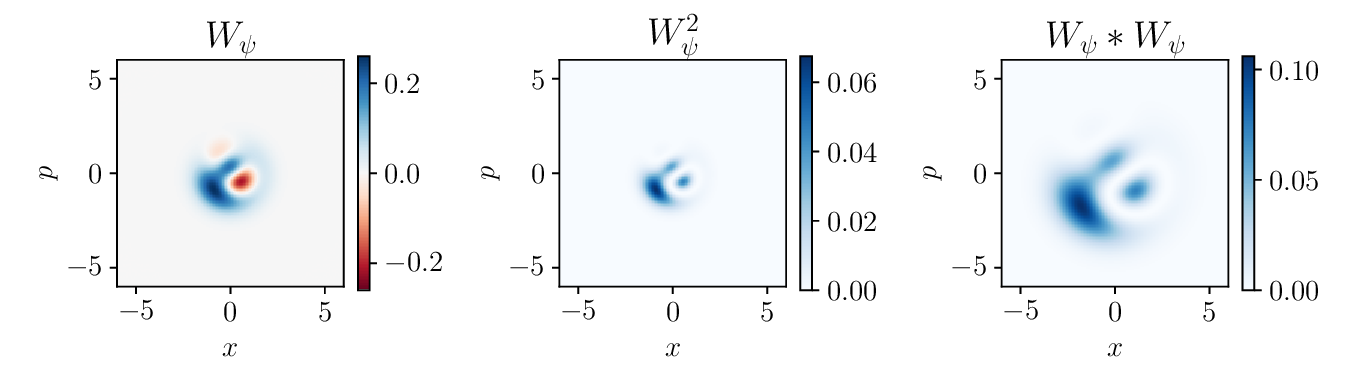}
    \caption{
        Illustration of the inner interference formula~\eqref{eq:inner_interference_formula} with the pure state $\ket{\psi}=\big(\ket{0}-\ket{1}+i\ket{2}\big)/\sqrt{3}$.
        From its Wigner function $W_\psi$, we compute $W_\psi^2$ and $W_\psi\ast W_\psi$, which are both non-negative. We observe that, up to a rescaling and a multiplicative constant, the two distributions are equal, confirming that $W_\psi\ast W_\psi=2\pi\, \mathcal{L}_{2}\big[W^2_\psi\big]$.
    }
    \label{fig:interference_formula}
\end{figure*}

In this section, we present an identity that relates the product of two cross Wigner functions to the convolution of two others. This formula, known as the interference formula, appears to have originated in the context of signal analysis, but has, to the best of our knowledge, not been exploited in quantum optics. It was first highlighted by Janssen~\cite{Janssen1979-es, Hlawatsch1984-yx}.

In what follows, we deal with operators of the type $\ket{\psi}\bra{\varphi}$, which are sometimes called interference or transition terms between the pure states $\ket{\psi}$ and $\ket{\varphi}$ (they appear when expressing the density operator for the superposition state $\ket{\psi}+\ket{\varphi}$).
These operators $\ket{\psi}\bra{\varphi}$ are non-Hermitian as soon as $\ket{\psi}\neq\ket{\varphi}$ and have trace $\Tr[\ket{\psi}\bra{\varphi}]=\bra{\varphi}\ket{\psi}$.
We denote their Weyl transform as
\begin{align}
    W_{\psi\varphi}(\alpha)
    =
    \frac{1}{\pi}
    \Big\langle\varphi\Big\vert\hat{\Pi}\left(\alpha/\sqrt{2}\right)\Big\vert\psi\Big\rangle.
    \label{eq:wigner_interference}
\end{align}
The function $W_{\psi\varphi}$ is called the cross Wigner function of $\ket{\psi}$ and $\ket{\varphi}$.
The particular case $\ket{\psi}\bra{\psi}$ yields the Wigner function of $\ket{\psi}$, which we note $W_\psi\vcentcolon=W_{\psi\psi}$.

As presented in Sec. \ref{sec:preliminaries}, the value of the Wigner function at the phase-space location $\alpha=x+ip$ is proportional to the expectation of a displaced parity $\hat{\Pi}(\alpha/\sqrt{2})$.
Observe now that the displaced parity operator $\hat{\Pi}(\alpha)$ performs a reflection with respect to the point $\alpha$ in phase space \cite{Grossmann1976-vk, Royer1977-he, Potocek2015-ds}.
We may thus commute (up to a change of sign) the displacement operator $\hat{D}(\alpha)$ and the reflection operator $\hat{\Pi}$ (with respect to the origin) in order to write $\hat{\Pi}(\alpha)=\hat{D}(\alpha)\hat{\Pi}\hat{D}(-\alpha)=\hat{D}(2\alpha)\hat{\Pi}$.
From this, it follows that if an operator $\hat{A}$ has the Weyl transform $A$, then we may associate to it the operator $\hat{B}_{\hat{A}}\vcentcolon=\hat{\Pi}(\alpha/\sqrt{2})\hat{A}\hat{\Pi}(\alpha/\sqrt{2})=\hat{D}(\sqrt{2}\, \alpha)\hat{\Pi} \hat{A}\hat{\Pi}\hat{D}^\dagger(\sqrt{2}\, \alpha)$ that has the Weyl transform $B(\beta)=A(2\alpha-\beta)$. This relation will be useful in the forthcoming development.

Let us now consider four pure states $\ket{a},\ket{b},\ket{c},\ket{d}$ and the different cross Wigner functions that they generate.
Using their definitions \eqref{eq:wigner_interference}, we may write:
\begin{align}
    W_{cb}(\alpha)W_{ad}(\alpha)
    = &
    \frac{1}{\pi}
    \bra{b}
    \hat{\Pi}\big(\alpha/\sqrt{2}\big)
    \ket{c}
    \
    \frac{1}{\pi}
    \bra{d}
    \hat{\Pi}\big(\alpha/\sqrt{2}\big)
    \ket{a}
    \nonumber
    \\[0.7em]
    = &
    \frac{1}{\pi^2}
    \mathrm{Tr}\Big[
    \ket{a}\bra{b} \underbrace{
        \hat{\Pi}\big(\alpha/\sqrt{2}\big)
        \ket{c}\bra{d}
        \hat{\Pi}\big(\alpha/\sqrt{2}\big) }_{ \Large {\hat B}_{\ket{c}\bra{d}} }
    \Big]
    \nonumber
    \\[0.7em]
    = &
    \frac{2}{\pi}\int
    W_{ab}(\beta) \,
    W_{cd}(2\alpha-\beta) \,
    \dd^2\beta
    \nonumber
    \\[0.7em]
    = &
    \frac{2}{\pi}\,
    \Big(W_{ab}\ast W_{cd}\Big)(2\alpha)
    \nonumber
    \\[0.8em]
    = &
    \frac{1}{2\pi}
    \
    \mathcal{L}_{1/2}
    \Big[
        W_{ab}\ast W_{cd}
        \Big](\alpha)
    \nonumber
\end{align}
where we have used the overlap formula as well as the Weyl transform $B(\beta)=A(2\alpha-\beta)$ of ${\hat B}_{\ket{c}\bra{d}}$ in the third equation. A little rearranging finally yields the so-called \textit{interference formula} \cite{Janssen1979-es, Hlawatsch1984-yx} (see also  \cite{Janssen1998-as}), namely
\begin{align}
    W_{ab}\ast W_{cd}=2\pi \,  \mathcal{L}_2[W_{cb}\cdot W_{ad}].
    \label{eq:general_interference_formula}
\end{align}
It holds for any pure states $\lbrace\ket{a},\ket{b},\ket{c},\ket{d}\rbrace$.
The special case of Eq.~\eqref{eq:general_interference_formula} with
$\ket{a}=\ket{b}\vcentcolon=\ket{\psi}$ and $\ket{c}=\ket{d}\vcentcolon=\ket{\varphi}$ is known as the \textit {outer interference formula}:
\begin{align}
    W_{\psi}\ast W_{\varphi}
    =
    2\pi \, \mathcal{L}_{2}
    \left[
        \abs{W_{\psi\varphi}}^2
        \right].
    \label{eq:outer_interference_formula}
\end{align}
Incidentally, Eq. \eqref{eq:outer_interference_formula} implies that the convolution of any two Wigner functions is always non-negative, which explains why BS states are Wigner positive. Indeed, the convolution in Eq. \eqref{eq:outer_interference_formula} is reminiscent to the convolution appearing in Eq. \eqref{eq:Wigner_fct_extreme_bs_state} when expressing the Wigner function of elementary BS states $\hat{\sigma}_\ast(\psi,\varphi)$; hence, such states must be Wigner positive and this property trivially extends to all BS states (i.e., convex mixtures of $\hat{\sigma}_\ast$'s).

Another interesting relation can be obtained from Eq.~\eqref{eq:general_interference_formula} by setting $\ket{a}=\ket{c}\vcentcolon=\ket{\psi}$ and $\ket{b}=\ket{d}\vcentcolon=\ket{\varphi}$, namely
\begin{align}
    W_{\psi\varphi}
    \ast
    W_{\psi\varphi}
    =
    2\pi \, \mathcal{L}_{2}\left[
        W_{\psi\varphi}^2
        \right],
    \label{eq:cross_interference_formula}
\end{align}
which we call the \textit{cross interference formula}. Notice here that, unlike in Eq. \eqref{eq:outer_interference_formula}, we have a square instead of a square modulus, so that both sides of the equation are complex-valued phase-space distributions.
Also, both sides of Eq. \eqref{eq:cross_interference_formula} have a quadratic dependence in $W_{\psi\varphi}$, so that it holds for any operator of the form $\lambda\ket{\psi}\bra{\varphi}$ where $\lambda\in\mathbb{C}$.
For $\ket{\psi}=\ket{\varphi}$, Eqs. \eqref{eq:outer_interference_formula} and \eqref{eq:cross_interference_formula} reduce to the same identity, namely
\begin{align}
    W_{\psi}\ast W_{\psi}
    =
    2\pi \, \mathcal{L}_{2}
    \left[
        W_\psi^2
        \right],
    \label{eq:inner_interference_formula}
\end{align}
which is called the \textit{inner interference formula} \cite{Janssen1982-op}.

At this point, let us take a moment to comment on the simple, yet surprising Eq. \eqref{eq:inner_interference_formula}.
It establishes a direct connection between the square of a pure Wigner function and its convolution with itself.
Note that this relates two terms of very different nature: evaluating $W_\psi^2(\alpha)$ only requires the knowledge of $W_\psi$ at a single phase-space location (in some sense, it is local); evaluating $(W_\psi\ast W_\psi)(\alpha)$ requires the knowledge of $W_{\psi}$ at all phase-space locations (in some sense, it is non-local).
Thus, Eq.~\eqref{eq:inner_interference_formula} is non-trivial and highly constrains the shape of $W_{\psi}$.
Note that  Eq. \eqref{eq:inner_interference_formula} cannot hold for mixed states.
Indeed, by integrating both sides we find $1=2\pi\int W^2(\alpha)\mathrm{d}^2\alpha$ which implies that the state as purity equal to $1$.
See Fig. \ref{fig:interference_formula} for an illustration.

The outer interference formula \eqref{eq:outer_interference_formula} is actually the only element that will be needed to perform the proofs in the next section, but we give more details on the interference formulas in Appendix~\ref{app-interference-formula} for completeness.

\section{Main results}
\label{sec:results}

We now come to the main result of this paper, namely that the entropy conjectures hold true for the family of BS states introduced in Sec. \ref{sec:beam-splitter-states}. More precisely, we show that all states $\hat{\sigma}\in\mathcal{B}$ satisfy Conj. \ref{conj:wig_entropy} as well as Conj. \ref{conj:wig_rényi_entropy} in the range $\alpha\in[1/2,\infty]$. Our proof is based on three ingredients: the characterization of BS states in phase space as in Sec. \ref{sec:beam-splitter-states}, the outer interference formula presented in Sec.~\ref{sec:interference_formula}, and bounds on the norms of cross Wigner functions that have been obtained by E. Lieb \cite{Lieb1990-ev}.

Hereafter, we recall the theorems of Ref. \cite{Lieb1990-ev} and adjust them to fit our conventions:
\begin{enumerate}
    \item[$(i)$]
          \quad
          $\Vert W_{\psi\varphi}\Vert_p\leq \Vert W_0\Vert_p$
          \qquad$\forall p\in(2,\infty]$
    \item[$(ii)$]
          \quad
          $\Vert W_{\psi\varphi}\Vert_p \geq \Vert W_0\Vert_p$
          \qquad$\forall p\in[1,2)$
    \item[$(iii)$]
          \quad
          $h(2\pi\abs{W_{\psi\varphi}}^2)\geq h(2\pi\abs{W_0}^2)$
\end{enumerate}
where $W_{\psi\varphi}$ is the cross Wigner function of $\ket{\psi},\ket{\varphi}$ and $W_0$ stands for the Wigner function of the vacuum. Note that the $2\pi$ factor in $(iii)$ ensures that the distribution is normalized.
Each of these inequalities is saturated if and only if $\ket{\psi},\ket{\varphi}$ are a matched Gaussian pair, which we define hereafter.

\begin{definition}[Matched Gaussian pair]
    Two Gaussian pure states $\ket{\gamma_1},\ket{\gamma_2}$ are a matched Gaussian pair iff they have equal covariance matrix.
\end{definition}

Equivalently, the Gaussian pure states $\ket{\gamma_1},\ket{\gamma_2}$ with wave functions $\gamma_1(x)\propto\mathrm{exp}(a_1 x^2+b_1 x)$ and $\gamma_2(x)\propto\mathrm{exp}(a_2 x^2+b_2 x)$ are a matched Gaussian pair iff $a_1=a_2$ ($a_1,a_2\in\mathbb{C}$ with $\mathrm{Re}~a_1<0$ and $\mathrm{Re}~a_2<0$).
Or, more simply, iff $\exists \alpha\in\mathbb{C}$ such that $\ket{\gamma_1}\propto\hat{D}(\alpha)\ket{\gamma_2}$.

A remarkable property of the cross Wigner function of a matched Gaussian pair is that its modulus coincides with the Wigner function of a Gaussian pure state.
More precisely, the cross Wigner function $W_{\gamma_1\gamma_2}$ of a matched Gaussian pair $\ket{\gamma_1},\ket{\gamma_2}$ satisfies:
\begin{align}
    \abs{W_{\gamma_1\gamma_2}(\alpha)}
    =
    W_0(\beta)
\end{align}
where $(\mathrm{Re}\beta,\mathrm{Im}\beta)^\intercal=\mathbf{S}\,(\mathrm{Re}\alpha,\mathrm{Im}\alpha)^\intercal+\mathbf{d}$ for some symplectic matrix $\mathbf{S}$ and displacement vector $\mathbf{d}$.
Recall that $W_0(\alpha)=\mathrm{exp}(-\abs{\alpha}^2)/\pi$ is the Wigner function of vacuum.
As a consequence, $\abs{W_{\gamma_1\gamma_2}}$ and $W_0$ are level-equivalent in the language of continuous majorization \cite{Van-Herstraeten2023-ww} and have equal $p$-norms, Shannon and Rényi entropies, hence they saturate inequalities $(i)-(iii)$.

%This means that $h_{\alpha}(W_{\gamma_1\gamma_2})=h_{\alpha}(W_0)$ for all matched Gaussian pair $\ket{\gamma_1},\ket{\gamma_2}$. 

\renewcommand{\arraystretch}{2.0}
\begin{table}[]
    \centering
    \resizebox{\columnwidth}{!}{%
        \begin{tabular}{|c|c|c|}
            \hline
            $W'$                         & $\Vert W'\Vert_p$                             & $h_\alpha(W')$                                  \\ \hhline{|=|=|=|}
            $\quad\mathcal{L}_s[W]\quad$ & $\quad s^{2\frac{1-p}{p}}\Vert W\Vert_p\quad$ & $\quad h_\alpha(W)+2\ln\abs{s}\quad$            \\ \hline
            $W^n$                        & $\Vert W\Vert_{np}^{n}$                       & $\frac{1-n\alpha}{1-\alpha}\; h_{n\alpha}(W)$   \\ \hline
            $c\cdot W$                   & $\abs{c}\cdot\Vert  W\Vert_p$                 & $h_\alpha(W)+\frac{\alpha\ln\abs{c}}{1-\alpha}$ \\ \hline
        \end{tabular}%
    }
    \caption{Three possible operations on a Wigner function:  rescaling by $s$ (first row), namely $W'=\mathcal{L}_s[W]$, exponentiation by a power $n$ (second row), namely $W'=W^n$, and multiplication by a constant $c$ (third row), namely $W'=c\cdot W$.
    In each case, the associated effect on the $p$-norm $\Vert W'\Vert_p$ and Rényi entropy $h_\alpha(W')$ is shown as a function of the $p$-norm and Rényi entropy of $W$.
    }
    \label{table:formulas-renyi-pnorms}
\end{table}

Interestingly, inequalities $(i)-(iii)$ can be combined into a single lower bound on the Rényi entropy of the distribution $2\pi\abs{W_{\psi\varphi}}^2$. As already mentioned, the factor $2\pi$ is needed to normalize the cross Wigner function so that the Rényi entropy tends towards the Shannon entropy as $\alpha\rightarrow 1$. Using Def.~\ref{def-Wigner-Renyi} and the fact that $h_\alpha(c\, W)=h_\alpha(W)+\alpha\ln \abs{c}/(1-\alpha)$ and $h_\alpha(W^n)=h_{n\alpha}(W)(1-n\alpha)/(1-\alpha)$, as summarized in Table~\ref{table:formulas-renyi-pnorms}, we can restate inequalities $(i)-(iii)$ as follows.

\begin{lemma}[Restatement of \cite{Lieb1990-ev}]
    Let $\ket{\psi},\ket{\varphi}$ be two pure states and $W_{\psi\varphi}$ be their cross Wigner function.
    The Rényi entropy of $\abs{W_{\psi\varphi}}^2$ is lower bounded as
    \begin{align*}
        h_\alpha\left(
        2\pi\abs{W_{\psi\varphi}}^2
        \right)
        \geq
        h_\alpha\left(
        2\pi\abs{W_0}^2\right)
        \qquad\forall\alpha\in[1/2,\infty]
    \end{align*}
    and equality is achieved iff $\ket{\psi},\ket{\varphi}$ are a matched Gaussian pair.
    \label{th:Lieb}
\end{lemma}

In some sense, Lemma \ref{th:Lieb} highlights that Gaussian pure states are minimum-uncertainty distributions among all cross Wigner functions.
Note that Lemma \ref{th:Lieb} covers the wide range $\alpha\in[1/2,\infty]$, notably including the Shannon entropy ($\alpha=1$) but excluding the range $\alpha\in(0,1/2)$.
% we will discuss the case $\alpha\in[0,1/2)$ later in this section.

In order to proceed, we will now exploit the results of Secs. \ref{sec:beam-splitter-states} and
\ref{sec:interference_formula}. Recalling expression \eqref{eq:Wigner_fct_extreme_bs_state} for the Wigner function of elementary BS states $\hat{\sigma}_\ast(\psi,\varphi)$, we have
\begin{align}
    W_{\hat{\sigma}_\ast(\psi,\varphi)}
    \ =\
    \mathcal{L}_{\frac{1}{\sqrt{2}}}
    \Big[W_{\psi}\ast W_{\varphi}\Big]
    \ =\
    2\pi\mathcal{L}_{\sqrt{2}}
    \Big[\abs{W_{\psi\varphi}}^2\Big]
\end{align}
where the second equality follows from the outer interference formula \eqref{eq:outer_interference_formula}.
From this, we obtain the following identity, which we express as a Lemma.

\begin{lemma}
    Let $\ket{\psi},\ket{\varphi}$ be two pure states. Their cross Wigner function $W_{\psi\varphi}$ satisfies
    \begin{align*}
        2\pi\abs{W_{\psi\varphi}}^2= \mathcal{L}_{\frac{1}{\sqrt{2}}}
        \Big[ W_{\hat{\sigma}_\ast(\psi,\varphi)} \Big]
    \end{align*}
    where $\hat{\sigma}_{\ast}(\psi,\varphi)$ is the elementary BS state produced by $\ket{\psi},\ket{\varphi}$ and  $W_{\hat{\sigma}_{\ast}(\psi,\varphi)}$ its associated Wigner function.
    \label{th:second-lemma}
\end{lemma}
We are now equipped to prove the main results of our paper.

\begin{theorem}[Wigner-Rényi entropy of BS states]
    The Wigner-Rényi entropy of any BS states is lower bounded by its value for the vacuum state, namely
    \begin{align}
        h_\alpha(W_{\hat{\sigma}})
        \geq
        h_\alpha(W_0)
        \qquad\forall\hat{\sigma}\in\mathcal{B},\quad\forall\alpha\in[1/2,\infty].
    \end{align}
    \label{theorem-Wigner-Renyi}
\end{theorem}

The proof combines Lemma~\ref{th:Lieb} and Lemma~\ref{th:second-lemma}. Noting that
$2\pi\abs{W_0}^2 =2\pi\abs{W_{0,0}}^2= \mathcal{L}_{1/\sqrt{2}}
    \big[ W_{\hat{\sigma}_\ast(0,0)} \big]=\mathcal{L}_{1/\sqrt{2}}
    \big[ W_0 \big]$,  we get
\begin{align}
    h_\alpha\left( \mathcal{L}_{\frac{1}{\sqrt{2}}}
    \Big[ W_{\hat{\sigma}_\ast(\psi,\varphi)} \Big] \right)
    \geq h_\alpha\left( \mathcal{L}_{\frac{1}{\sqrt{2}}}
    \Big[ W_0 \Big] \right) \quad\forall\alpha\in[1/2,\infty].
\end{align}
Then, using the straightforward derivations presented in Table \ref{table:formulas-renyi-pnorms}, we get the fundamental inequality
\begin{align}
    h_\alpha\big( W_{\hat{\sigma}_\ast(\psi,\varphi)} \big)
    \geq
    h_\alpha\big(W_0\big)
    \quad\forall\alpha\in[1/2,\infty].
    \label{eq-fundamental-ineq}
\end{align}
This proves that all elementary BS state $\hat{\sigma}_\ast(\psi,\varphi)\in\mathcal{B}_\ast$ satisfy the Wigner-Rényi entropy conjecture (Conj. \ref{conj:wig_rényi_entropy}) for $\alpha\in[1/2,\infty]$.

Remember now that the Wigner function of any BS state $\hat{\sigma}$ in $\mathcal{B}$ can be written as a convex mixture of the Wigner functions of elementary BS states in $\mathcal{B}_\ast$, namely $W_{\hat{\sigma}}=\sum_i p_i W_{\hat{\sigma}_\ast(\psi_i,\varphi_i)} $. Since the Rényi entropy is concave in the range $\alpha\in[0,1]$, namely $h_\alpha(\sum_i p_i W_i)\ge \sum_i p_i \, h_\alpha(W_i)$, Eq. \eqref{eq-fundamental-ineq} trivially extends to all BS states $\hat{\sigma}\in\mathcal{B}$ provided $\alpha\in[1/2, 1]$. Furthermore, it is known that the Rényi entropy remains quasi-concave in the range $\alpha\in(1, \infty]$, namely $h_\alpha(\sum_i p_i W_i)\ge \min_i h_\alpha(W_i)$. While it is weaker than concavity, this property is sufficient to extend Eq. \eqref{eq-fundamental-ineq} to all BS states in the range $\alpha\in(1, \infty]$ since the right-hand side term of Eq.~\eqref{eq-fundamental-ineq} is a constant (independent of $i$). This concludes the proof of Theorem~\ref{theorem-Wigner-Renyi} (which validates Conj.~\ref{conj:wig_rényi_entropy} for BS states albeit in the range $\alpha\in[1/2,\infty]$). In particular, this is true for $\alpha=1$, which brings us to our second theorem (which validates Conj.~\ref{conj:wig_entropy} for BS states).

%that is, $\smash{W=\sum_i p_i \mathcal{L}_{1/\sqrt{2}}[W_{\psi_i}\ast W_{\varphi_i}]=2\pi\sum_i p_i\mathcal{L}_{\sqrt{2}}[\abs{W_{\psi_i\varphi_i}}^2]}$. This decomposition makes apparent that BS states have non-negative Wigner functions.

%we have $h_\alpha(W_{\hat{\sigma}})\geq h_\alpha(W_0)$ 

\begin{theorem}[Wigner entropy of BS states]
    The Wigner entropy of any BS state is lower bounded by its value for the vacuum state, namely
    \begin{align}
        h(W_{\hat{\sigma}})
        \geq
        \ln\pi+1
        \qquad
        \forall\hat{\sigma}\in\mathcal{B}
    \end{align}
    \label{theorem-Wigner}
\end{theorem}

\begin{figure}
    \centering
    \includegraphics[width=0.95\linewidth]{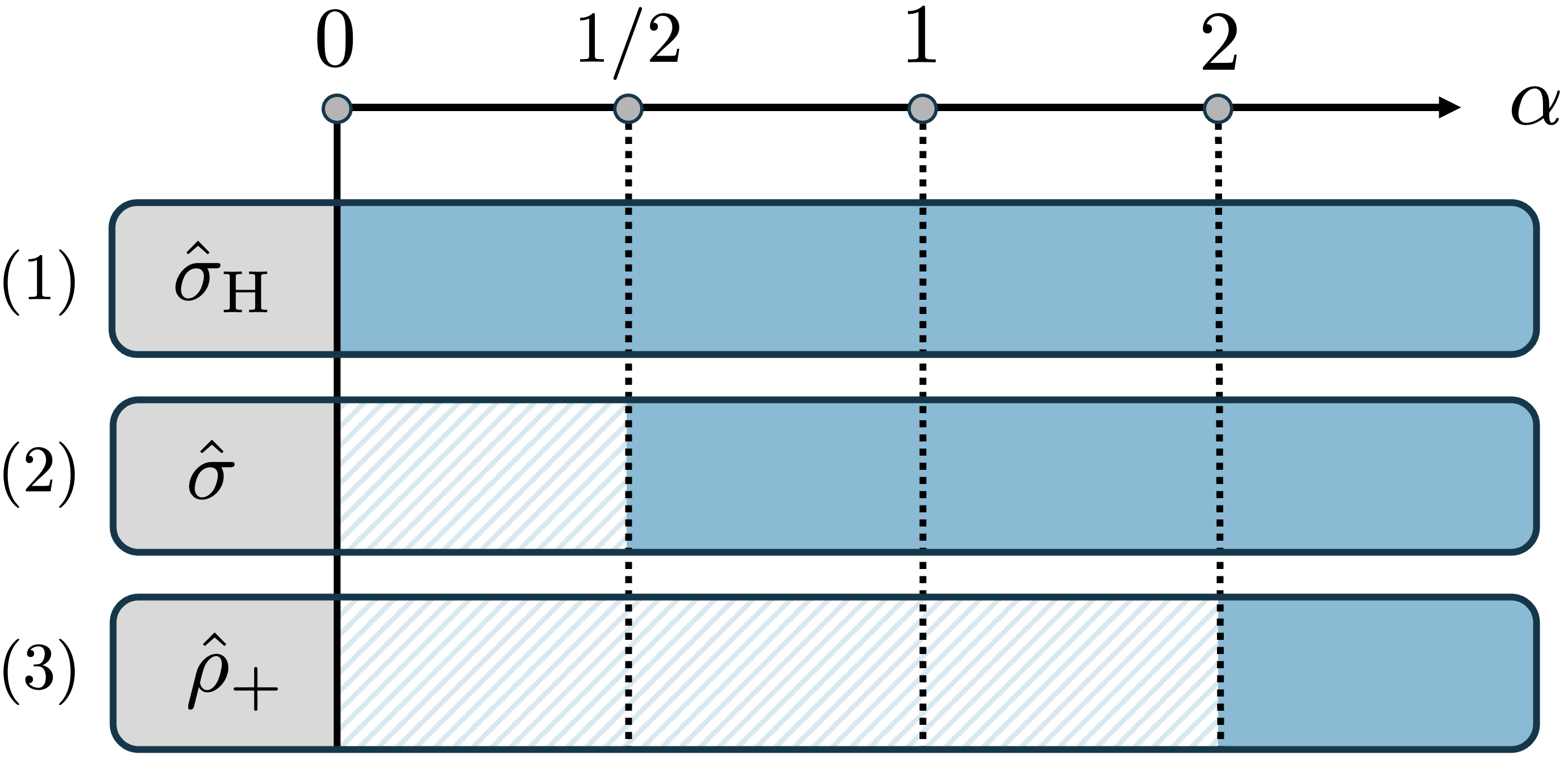}
    \caption{Schematic representation of the proven validity regions of the Wigner-Rényi entropy conjecture as a function of the parameter $\alpha$ for different classes of states.
        (1) Husimi BS states $\hat{\sigma}_\mathrm{H} \vcentcolon= \hat{\sigma}(\hat \rho,0)$, $\forall \hat\rho$, see Ref.~\cite{Lieb2014-qx}; (2) BS states $\hat{\sigma}\in\mathcal{B}$, see Theorem \ref{theorem-Wigner-Renyi}; (3) Wigner-positive state $\hat{\rho}_+\in\mathcal{W}_+$, see Ref.~\cite{Dias2023-qr}.
    }
    \label{fig:wigner-renyi-conjecture-ppt}
\end{figure}

In summary, Theorem \ref{theorem-Wigner-Renyi} proves the Wigner-Rényi entropy conjecture for BS states with $\alpha\in[1/2,\infty]$. Unfortunately, the proof does not extend to $\alpha<1/2$ because the fundamental inequality \eqref{eq-fundamental-ineq} is not proven in this range (which comes in turn from the proven validity range of Lemma~\ref{th:Lieb}). Note that Theorem \ref{theorem-Wigner-Renyi} actually extends to the entire range $\alpha\in \, (0,\infty]$ for the special class of Husimi BS states (i.e., states whose Wigner function is a valid Husimi Q function) as a consequence of Ref.~\cite{Lieb2014-qx}.
Further, Theorem \ref{theorem-Wigner} corresponds to the special case $\alpha =1$ and proves the Wigner entropy conjecture for BS states since the Shannon entropy is concave. Finally, remember that Ref.~\cite{Dias2023-qr} proves that, for any Wigner-positive state, the Wigner-Rényi entropy conjecture is satisfied for $\alpha\geq 2$. The proven validity regions of the conjecture for different set of states is illustrated in Fig.~\ref{fig:wigner-renyi-conjecture-ppt}.

%Then, using the concavity of Rényi entropy for $\alpha\in[0, 1]$ it also proves the Wigner-Rényi entropy conjecture .

\section{Discussion and conclusion}
\label{sec:conclusion}

The Wigner-entropy conjecture suggests the existence of a fundamental bound on the uncertainty of all quantum states that admit a classical description in their Wigner representation (i.e., Wigner-positive states).
Recently, the conjecture has received increasing attention, with several partial results hinting to its validity.
In this work, we prove that the conjecture is true for the large family of Wigner-positive states known as beam-splitter states (Theorem~\ref{theorem-Wigner}).
Further, we show that these states even satisfy the extended Wigner-Rényi entropy conjecture in some rather large range of the Rényi parameter $\alpha$ (Theorem~\ref{theorem-Wigner-Renyi}).

In Sec.~\ref{sec:beam-splitter-states}, we give an overview of beam-splitter states, whose convex set is noted $\mathcal{B}$.
We highlight their relation with the set of separable states, and identify $\mathcal{B}_\ast$ as the subset of beam-splitter states that originate from pure separable input states (we refer to these states in $\mathcal{B}_{\ast}$ as \textit{elementary} beam-splitter states). Since the whole set~$\mathcal{B}$ can be obtained from convex mixtures of states of $\mathcal{B}_{\ast}$, lower bounds on the Wigner entropy of states of $\mathcal{B}_{\ast}$ automatically extend to all states of $\mathcal{B}$ as a consequence of the concavity of Shannon entropy (the situation is slightly more complicated for the Rényi entropy).

The core ingredient to our result is the \textit{interference formula}, which relates a convolution of cross Wigner functions to a product of (conjugate) cross Wigner functions.
We devote Sec. \ref{sec:interference_formula} to this formula, which we present in a quantum-optical notation.
In particular, we focus on a specific instance of the relation known as the \textit{outer interference formula}, which relates the convolution of two pure Wigner functions to the squared modulus of their cross Wigner function.
Since a beam splitter performs a convolution in phase space, the outer interference formula allows us to interpret the Wigner function of any state in $\mathcal{B}_\ast$ as the squared modulus of a cross Wigner function (or a convex mixture thereof for states in $\mathcal{B}$).

Then, in Sec.~\ref{sec:results}, we build upon existing bounds on the $p$-norms of cross Wigner functions to show that the Wigner-Rényi entropy of parameter $\alpha$ of states in $\mathcal{B}_{\ast}$ satisfies the conjectured lower bound in the regime $\alpha\geq 1/2$. Using a (quasi-)concavity argument, we then extend this proof to all states in $\mathcal{B}$.
In particular, the case $\alpha=1$ implies that all beam-splitter states satisfy the Wigner entropy conjecture.
Incidentally, when applied to Husimi beam-splitter states, i.e., beam-splitter states of the form $\hat{\sigma}(\hat{\rho},0)$, our results yield an alternative proof of the Wehrl entropy conjecture, long proved in Refs. \cite{wehrl1979relation, Lieb1978-gy}.

As stressed earlier, Lemma~\ref{th:Lieb}, and by extension Theorem~\ref{theorem-Wigner-Renyi}, do not apply to the regime $\alpha\in(0,1/2)$.
Nevertheless, we still expect the lower bound to be valid in that regime, i.e., we expect $h_\alpha(2\pi\abs{W_{\psi\varphi}}^2)\geq h_\alpha(2\pi\abs{W_0}^2)$ to hold for all $\alpha > 0$.
This stronger statement would in fact be implied by the validity of the Wigner-Rényi conjecture (Conj. \ref{conj:wig_rényi_entropy}). We may wonder why is Lemma~\ref{th:Lieb} limited to $\alpha\geq 1/2$?
The proof presented in Ref.~\cite{Lieb1990-ev} relies on inequalities using conjugated exponents $p,q$ such that $1/p+1/q=1$.
One of the exponents being lower than $1$ implies the other to be negative, but the $p$-norm is well defined for non-negative $p$ values only. For this reason, the proof of Ref.~\cite{Lieb1990-ev} is limited to $p$-norms of parameter $p\geq 1$, which in Lemma~\ref{th:Lieb} translates to $\alpha\geq 1/2$.

\begin{figure}
    \centering
    \includegraphics[width=\linewidth]{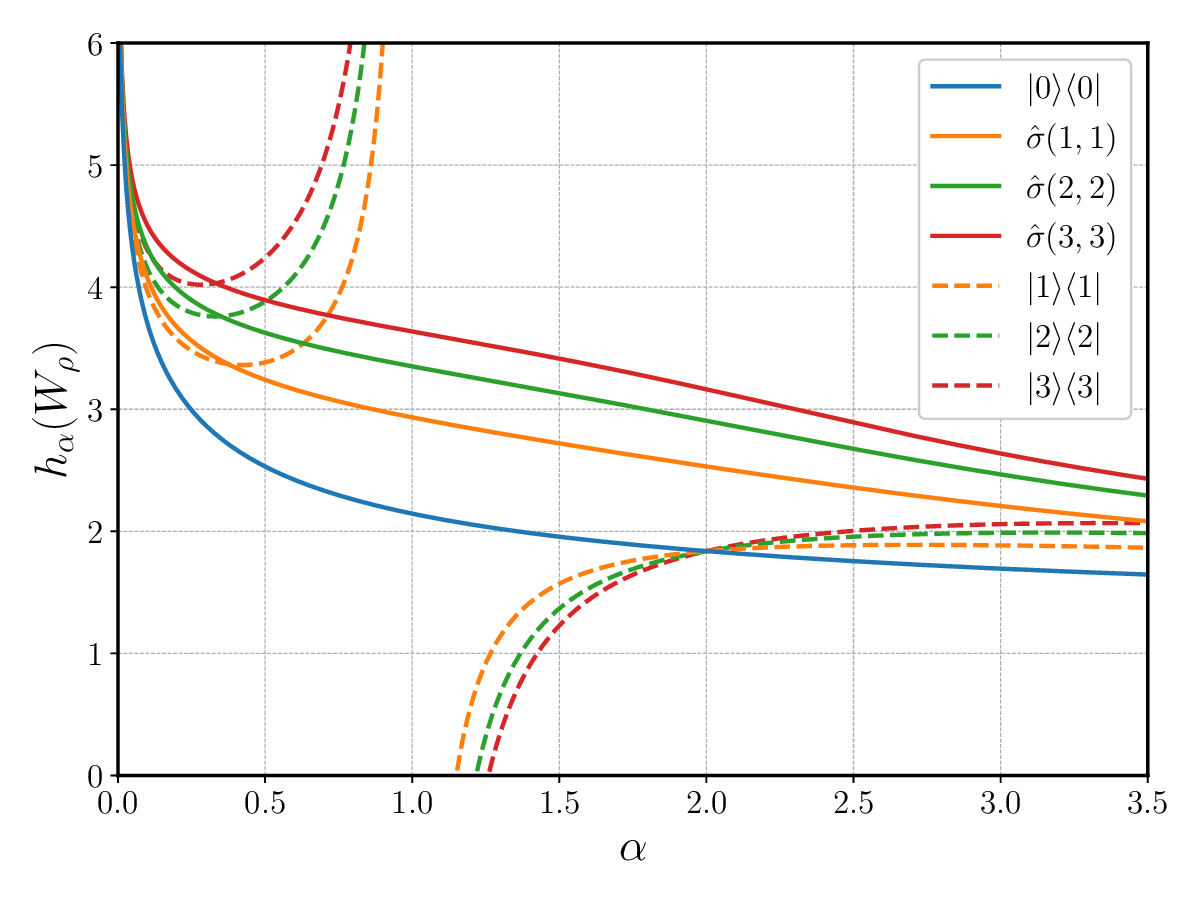}
    \caption{
        Plot of the Wigner-Rényi entropy of several quantum states as a function of $\alpha$.
        We consider the Wigner-positive states $\hat{\sigma}(0,0)=\ket{0}\bra{0}$, $\hat{\sigma}(1,1)$, $\hat{\sigma}(2,2)$, $\hat{\sigma}(3,3)$ and observe that their Wigner-Rényi entropy is lower-bounded by $h_\alpha(W_0)$ in accordance with Conjecture \ref{conj:wig_rényi_entropy}.
        Then, we consider the Wigner-negative pure states $\ket{1},\ket{2},\ket{3}$ and observe that their Wigner-Rényi entropy is lower bounded by $h_\alpha(W_0)$ for $\alpha\in(0,1)\cup[2,\infty]$ and upper bounded by $h_\alpha(W_0)$ for $\alpha\in(1,2]$, the cross-over thus taking place at  $\alpha=2$. In the limit $\alpha\rightarrow 1$, the Wigner-Rényi entropy of Wigner-negative states diverges.
    }
    \label{fig:wigner-renyi-plot}
\end{figure}

Interestingly, Lemma~\ref{th:Lieb} can also be used to derive inequalities for the Wigner-Rényi entropy of pure states.
Note here that, in contrast to beam-splitter states, the Wigner function of a pure state generally takes negative values, implying $\Vert W_\psi\Vert_1>1$.
This means that $h_{\alpha}(W_\psi)$ diverges in the limit $\alpha\rightarrow 1$, as can indeed been seen in Fig.~\ref{fig:wigner-renyi-plot}.
For any pure state $\ket{\psi}$, it follows from Lemma~\ref{th:Lieb} and Table~\ref{table:formulas-renyi-pnorms} that $h_{\alpha}(W_{\psi})\geq h_{\alpha}(W_0)$ when $\alpha\geq 2$, while the inequality is reversed when $1<\alpha\leq 2$ (see Fig.~\ref{fig:wigner-renyi-plot}).
Viewing the Wigner-Rényi entropy as a measure of uncertainty, this underlines the ambivalent role of Gaussian pure states.
Indeed, among the set of pure states, Gaussian states maximize the Wigner-Rényi entropy in the regime $\alpha\in(1,2]$, but minimize it in the regime $\alpha\in[2,\infty]$.
Note that we also expect Gaussian pure states to minimize $h_\alpha$ among pure states when $\alpha\in(0,1)$ although this remains to be proven (such a proof would in fact imply the validity of Lemma~\ref{th:Lieb} $\forall\alpha\geq 0$).

To date, the set $\mathcal{B}$ of beam-splitter states is arguably the largest known family of Wigner-positive states.
It benefits from a simple quantum-optical implementation from its definition, while it encompasses a great diversity of states.
Indeed, the set $\mathcal{B}$ includes varieties of quantum states with different shapes and symmetries (phase invariance or not) and contains, for example, the whole family of Gaussian states and passive states. Our results thus represent a significant advance towards a proof of the Wigner-entropy conjecture, though they do not make a full proof since there exist Wigner-positive states that are not beam-splitter states \cite{Van_Herstraeten2021-nj}.
Because of this hassle, the Wigner-entropy conjecture and its Wigner-Rényi extension remain open as of today.

At this point, we expect that further advance in proving the Wigner-entropy conjecture may come from a finer characterization of the set of Wigner-positive states.
In particular, the set of states that are Wigner-positive but lie outside the beam-splitter set $\mathcal{B}$ deserve further attention since this is precisely where the Wigner-entropy conjecture remains open. Another significant extension of this work could be to attempt proving the majorization conjecture~\cite{Van-Herstraeten2023-ww}, even restricted to BS states, which would automatically validate the Wigner-Rényi entropy conjecture for those states in the full range $\alpha\in(0,\infty]$. Yet another potential avenue could be to investigate the intriguing relationship between the Wigner-entropy conjecture and the minimum output entropy conjecture \cite{PhysRevA.70.032315} (now proved in \cite{,Giovannetti2014}), namely that coherent states minimize the output von Neumann entropy of all phase-insensitive Gaussian channels.

%{\color{red}Beyond Wigner-positive states, our result proves that any phase-space distribution with a PSD kernel  and covariant under displacement is subject to a lower-bound on its entropy.}

%\newpage
%\noindent

\vspace{-1em}
\acknowledgments

We thank Michael Jabbour for insightful discussions.
ZVH thanks Jack Davis and Ulysse Chabaud for fruitful exchanges.
NJC is grateful to the James C. Wyant College of Optical Sciences for hospitality during his sabbatical leave in the autumn 2022, when a large part of this work was carried out.
ZVH acknowledges funding from the European Union’s Horizon Europe Framework Programme (EIC Pathfinder Challenge project Veriqub) under Grant Agreement No.~101114899. NJC acknowledges support from the Fonds de la Recherche Scientifique – FNRS under the Excellence of Science project CHEQS.

%NJC acknowledges support from the European Union and the Fonds de la Recherche Scientifique – FNRS under project ShoQC within ERA-NET Cofund in Quantum Technologies (QuantERA) program.

\appendix

\vspace{-0.5em}

\section{Extreme beam-splitter states}
\label{app-extreme-BS}

\vspace{-0.5em}
As it appears from Definition \ref{def:bs_states}, the set $\mathcal{B}$ of BS states is intimately linked to the set of separable states, which we note $\mathcal{S}$. In fact, Eq.~\eqref{eq:rho_sep_bs_state} defines a linear map $\mathcal{S}\rightarrow\mathcal{B}$ and thus $\mathcal{B}$ inherits the convexity of $\mathcal{S}$. Furthermore, just as $\mathcal{S}$ is the convex hull of the set of pure product states $\ket{\psi}\ket{\varphi}$, which we note $\mathcal{S}_\ast$, we see that $\mathcal{B}$ is the convex hull of the set $\mathcal{B}_\ast$ of states $\hat{\sigma}_\ast(\psi,\varphi)$. As a consequence, all extreme states of $\mathcal{B}$ must be in $\mathcal{B}_\ast$. It is tempting to extrapolate that all states in $\mathcal{B}_\ast$ are  extreme states of $\mathcal{B}$, just as all states in $\mathcal{S}_\ast$ are  extreme states of $\mathcal{S}$, but this is actually wrong.

Indeed, the analogy is not perfect because the map $\mathcal{S}\rightarrow\mathcal{B}$ is \textit{surjective}: two different separable states may yield the same BS state.
For example, a BS state is invariant under exchange of its two inputs, $\hat{\sigma}(\hat{\rho}_1,\hat{\rho}_2)=\hat{\sigma}(\hat{\rho}_2,\hat{\rho}_1)$, which results from the commutativity of convolution. The surjectivity of the map $\mathcal{S}\rightarrow\mathcal{B}$ implies that the preimage of extreme points of $\mathcal{B}$ are extreme points of $\mathcal{S}$, but that the image of extreme points of $\mathcal{S}$ are not necessarily extreme points of $\mathcal{B}$.

Let us show an example of a state  $\hat{\sigma}_\ast(\psi,\varphi)$ which is not an extreme point of $\mathcal{B}$.
Consider the squeezing operator $\hat{S}(r)\vcentcolon=\mathrm{exp}((r\hat{a}^{\dagger 2}-r^\ast\hat{a}^{2})/2)$ and take $\ket{\psi}=\hat{S}(r)\ket{0}$ and $\ket{\varphi}=\hat{S}(-r)\ket{0}$ to be two squeezed states with orthogonal squeezing.
The beam splitter produces an (entangled) two-mode squeezed vacuum state, so that the resulting BS state is a thermal state. Since a thermal state is a mixture of coherent states (which are BS states), $\hat{\sigma}_\ast(\psi,\varphi)$ is not an extreme point of $\mathcal{B}$.
However, we believe that this situation is due to the infinite support of the squeezed states $\ket{\psi}$ and  $\ket{\varphi}$. In contrast, we expect that if $\ket{\psi}$  and $\ket{\varphi}$ have a finite Fock support, all BS states $\hat{\sigma}_\ast(\psi,\varphi)$ are extreme points of $\mathcal{B}$, so that $\mathcal{B}_\ast$ would then actually coincide with the set of extreme points of $\mathcal{B}$.
This question is worth future investigation, but has no impact on the results of this paper.

\vspace{-0.5em}

\section{Remarks on the interference formula}
\label{app-interference-formula}

\vspace{-0.5em}

Let us first comment on the factor $2\pi$ appearing in the inner interference formula, Eq. \eqref{eq:inner_interference_formula}. Remember that rescaling a Wigner function is equivalent to changing the units of the Planck constant $\hbar$.
Introducing the notation $W_{\psi,s}\vcentcolon=\mathcal{L}_s[W_\psi]$, we can write the inner interference formula \eqref{eq:inner_interference_formula} as
\begin{align}
    W_{\psi,s}
    \ast
    W_{\psi,s}
    =
    2\pi s^2\;
    \mathcal{L}_2\left[
    W^2_{\psi,s}.
    \right]
    \label{eq:interference_formula_rescaling}
\end{align}
The above equation shows that the multiplicative constant can be tuned by changing the value of $\hbar$; however the interference formula will always contain a rescaling operator $\mathcal{L}_2$, regardless of the value of $\hbar$.
Interestingly, choosing a rescaling $s=1/\sqrt{2\pi}$ in Eq. \eqref{eq:interference_formula_rescaling} (which corresponds to setting $\hbar=1/2\pi$ and $h=1$) precisely cancels the multiplicative constant and gives $W\ast W=\mathcal{L}_2[W^2]$.

In fact, we can also write a unit-dependant version of Eq. \eqref{eq:inner_interference_formula}.
Simply observe that the $\hbar$-dependant Wigner function is $W_\Psi(\alpha)\vcentcolon=W_\psi(\alpha/\sqrt{\hbar})/\hbar=\mathcal{L}_{\sqrt{\hbar}}[W_\psi](\alpha)$, and has units $[W_\Psi]=1/\hbar$.
Its argument has units $[\alpha]=\sqrt{\hbar}$.
Using Eq. \eqref{eq:interference_formula_rescaling}, we have then:
\begin{align}
    W_{\Psi}
    \ast
    W_{\Psi}
    =
    h\;
    \mathcal{L}_2\left[W_\Psi^2\right].
    \label{eq:inner_interference_formula_units}
\end{align}
The differential elements in the convolution have units $[\dd^2\alpha]=\hbar$, so that both sides of Eq. \eqref{eq:inner_interference_formula_units} have the units of $1/\hbar$.
Note that Eq. \eqref{eq:inner_interference_formula_units} can easily be generalized to write a $\hbar$-dependant version of the general interference formula, Eq. \eqref{eq:general_interference_formula}.

Le us now consider the special case of the inner interference formula, Eq. \eqref{eq:inner_interference_formula}, when considering Gaussian pure states. For any Gaussian distribution $G$ (Wigner function or not), we have $G\ast G=\mathcal{L}_{\sqrt{2}}\left[G\right]$.
Thus, when $W_\gamma$ is the Wigner function of a Gaussian pure state $\ket{\gamma}$,  Eq. \eqref{eq:inner_interference_formula} yields
\begin{align}
    W_\gamma= \mathcal{L}_{1/\sqrt{2}}[W_\gamma\ast W_\gamma]
    =
    2\pi \, \mathcal{L}_{\sqrt{2}}[W^2_\gamma]
    .
\end{align}
While the identity $W_\gamma=\mathcal{L}_{1/\sqrt{2}}[W_\gamma\ast W_\gamma]$ holds for any Gaussian distribution (even when the variance violates the Heisenberg uncertainty relation), the identity $W_\gamma= 2\pi \, \mathcal{L}_{\sqrt{2}}[W^2_\gamma]$ selects only the ones that precisely match the minimum uncertainty allowed by quantum mechanics (hence, the Gaussian pure states). Beyond Gaussian states, it is unknown whether Eq. \eqref{eq:inner_interference_formula} may also be sufficient to single out pure states.

This brings us to the following question about the inner interference formula. As previously noted, this fundamental relation obeyed by the Wigner function of all pure states establishes a connection between two unexpectedly related terms, imposing thus a stringent constraint on the shape of pure Wigner functions.
Observe that, as it is a strict equality between two functions of $\alpha$, it encapsulates an infinite number of equalities (one for each value of $\alpha$).
With these considerations in mind, we make the suggestion that the inner interference formula might in fact be a sufficient condition for the physicality of a phase-space distribution.
More specifically, we pose the following question:
\newline
%\textbf{Open question.}
\textit{
Let $F(\alpha)$ be a real-valued normalized distribution that satisfies $F\ast F=2\pi\mathcal{L}_2[F^2]$.
Is $F(\alpha)$ always an acceptable Wigner function? That is, does it always correspond to a positive-semidefinite operator?
}
\newline
If the answer is positive, then it is a first step towards a criterion for physicality that is embodied in the phase-space formalism. If the answer is negative, then it unveils a family of quasi-states that share a strong symmetry together with quantum pure states.
In both cases, further investigating the question is valuable, though again it has no consequence on the results of this paper.

%Relating the Wigner entropy of $\hat{\sigma}(\psi,\varphi)$ to the entropy of $\ket{\psi}$ and $\ket{\varphi}$.

%For a given $\ket{\psi}$, what is the $\ket{\varphi}$ that minimizes the Wigner entropy of $\hat{\sigma}(\psi,\varphi)$.

Finally, on a different matter, let us stress a remarkable property of the general interference formula \eqref{eq:general_interference_formula}, namely its invariance under the Fourier transformation.
This can be expected from the duality of multiplication and convolution when going to Fourier space.
In light of this, we define the Fourier transform as $\mathcal{F}[W](\beta)=\tilde{W}(\beta)$ with:
\begin{align}
    \tilde{W}(\beta)=\int W(\alpha)\exp(-i(\alpha^\ast\beta+\alpha\beta^\ast))\dd^2\alpha/2\pi.
\end{align}
The above expression is the usual Fourier transform; the symplectic Fourier transform is obtained as $\mathcal{F}_{\mathrm{s}}[W](\beta)=\mathcal{F}[W](i\beta)$.
Note that the Fourier transform of a Wigner function is in general complex-valued.
Then, we compute $\mathcal{F}[W_{ab}\ast W_{cd}]=2\pi\,\tilde{W}_{ab}\cdot\tilde{W}_{cd}$ and $\mathcal{F}[2\pi\mathcal{L}_2[W_{cb}\cdot W_{ad}]]=(1/4)\,\mathcal{L}_{1/2}[\tilde{W}_{cb}\ast\tilde{W}_{ad}]$.
After rearrangement, this yields:
\begin{align}
    \tilde{W}_{cd}\ast\tilde{W}_{ad}=8\pi\mathcal{L}_2[\tilde{W}_{ab}\cdot\tilde{W}_{cd}],
\end{align}
which is (up to a multiplicative constant on the RHS) the same relation as Eq. \eqref{eq:general_interference_formula} but this time expressed in Fourier space.
As we have just highlighted, the multiplicative constant is irrelevant since it can be suppressed with a change of units.
If we had chosen conventions such that $W_0$ is proportional to its own Fourier transform (i.e. setting $\hbar=2$), the interference formula would read exactly the same in phase space and in its dual under a Fourier transform.

\vspace{-1em}

\bibliography{BS-states}

\begin{thebibliography}{55}%
\makeatletter
\providecommand \@ifxundefined [1]{%
 \@ifx{#1\undefined}
}%
\providecommand \@ifnum [1]{%
 \ifnum #1\expandafter \@firstoftwo
 \else \expandafter \@secondoftwo
 \fi
}%
\providecommand \@ifx [1]{%
 \ifx #1\expandafter \@firstoftwo
 \else \expandafter \@secondoftwo
 \fi
}%
\providecommand \natexlab [1]{#1}%
\providecommand \enquote  [1]{``#1''}%
\providecommand \bibnamefont  [1]{#1}%
\providecommand \bibfnamefont [1]{#1}%
\providecommand \citenamefont [1]{#1}%
\providecommand \href@noop [0]{\@secondoftwo}%
\providecommand \href [0]{\begingroup \@sanitize@url \@href}%
\providecommand \@href[1]{\@@startlink{#1}\@@href}%
\providecommand \@@href[1]{\endgroup#1\@@endlink}%
\providecommand \@sanitize@url [0]{\catcode `\\12\catcode `\$12\catcode `\&12\catcode `\#12\catcode `\^12\catcode `\_12\catcode `\%12\relax}%
\providecommand \@@startlink[1]{}%
\providecommand \@@endlink[0]{}%
\providecommand \url  [0]{\begingroup\@sanitize@url \@url }%
\providecommand \@url [1]{\endgroup\@href {#1}{\urlprefix }}%
\providecommand \urlprefix  [0]{URL }%
\providecommand \Eprint [0]{\href }%
\providecommand \doibase [0]{https://doi.org/}%
\providecommand \selectlanguage [0]{\@gobble}%
\providecommand \bibinfo  [0]{\@secondoftwo}%
\providecommand \bibfield  [0]{\@secondoftwo}%
\providecommand \translation [1]{[#1]}%
\providecommand \BibitemOpen [0]{}%
\providecommand \bibitemStop [0]{}%
\providecommand \bibitemNoStop [0]{.\EOS\space}%
\providecommand \EOS [0]{\spacefactor3000\relax}%
\providecommand \BibitemShut  [1]{\csname bibitem#1\endcsname}%
\let\auto@bib@innerbib\@empty
%</preamble>
\bibitem [{\citenamefont {Albarelli}\ \emph {et~al.}(2018)\citenamefont {Albarelli}, \citenamefont {Genoni}, \citenamefont {Paris},\ and\ \citenamefont {Ferraro}}]{Albarelli2018-ei}%
  \BibitemOpen
  \bibfield  {author} {\bibinfo {author} {\bibfnamefont {F.}~\bibnamefont {Albarelli}}, \bibinfo {author} {\bibfnamefont {M.~G.}\ \bibnamefont {Genoni}}, \bibinfo {author} {\bibfnamefont {M.~G.~A.}\ \bibnamefont {Paris}},\ and\ \bibinfo {author} {\bibfnamefont {A.}~\bibnamefont {Ferraro}},\ }\bibfield  {title} {\bibinfo {title} {Resource theory of quantum non-{G}aussianity and {W}igner negativity},\ }\href {https://doi.org/10.1103/PhysRevA.98.052350} {\bibfield  {journal} {\bibinfo  {journal} {Phys. Rev. A}\ }\textbf {\bibinfo {volume} {98}},\ \bibinfo {pages} {052350} (\bibinfo {year} {2018})}\BibitemShut {NoStop}%
\bibitem [{\citenamefont {Chabaud}\ \emph {et~al.}(2021)\citenamefont {Chabaud}, \citenamefont {Emeriau},\ and\ \citenamefont {Grosshans}}]{Chabaud2021-tq}%
  \BibitemOpen
  \bibfield  {author} {\bibinfo {author} {\bibfnamefont {U.}~\bibnamefont {Chabaud}}, \bibinfo {author} {\bibfnamefont {P.-E.}\ \bibnamefont {Emeriau}},\ and\ \bibinfo {author} {\bibfnamefont {F.}~\bibnamefont {Grosshans}},\ }\bibfield  {title} {\bibinfo {title} {Witnessing {W}igner negativity},\ }\href {https://doi.org/10.22331/q-2021-06-08-471} {\bibfield  {journal} {\bibinfo  {journal} {Quantum}\ }\textbf {\bibinfo {volume} {5}},\ \bibinfo {pages} {471} (\bibinfo {year} {2021})}\BibitemShut {NoStop}%
\bibitem [{\citenamefont {Lee}(1991)}]{Lee1991-io}%
  \BibitemOpen
  \bibfield  {author} {\bibinfo {author} {\bibfnamefont {C.~T.}\ \bibnamefont {Lee}},\ }\bibfield  {title} {\bibinfo {title} {Measure of the nonclassicality of nonclassical states},\ }\href {https://doi.org/10.1103/physreva.44.r2775} {\bibfield  {journal} {\bibinfo  {journal} {Phys. Rev. A}\ }\textbf {\bibinfo {volume} {44}},\ \bibinfo {pages} {R2775} (\bibinfo {year} {1991})}\BibitemShut {NoStop}%
\bibitem [{\citenamefont {Lütkenhaus}\ and\ \citenamefont {Barnett}(1995)}]{Lutkenhaus1995-gr}%
  \BibitemOpen
  \bibfield  {author} {\bibinfo {author} {\bibfnamefont {N.}~\bibnamefont {Lütkenhaus}}\ and\ \bibinfo {author} {\bibfnamefont {S.~M.}\ \bibnamefont {Barnett}},\ }\bibfield  {title} {\bibinfo {title} {Nonclassical effects in phase space},\ }\href {https://doi.org/10.1103/physreva.51.3340} {\bibfield  {journal} {\bibinfo  {journal} {Phys. Rev. A}\ }\textbf {\bibinfo {volume} {51}},\ \bibinfo {pages} {3340} (\bibinfo {year} {1995})}\BibitemShut {NoStop}%
\bibitem [{\citenamefont {De~Bièvre}\ \emph {et~al.}(2019)\citenamefont {De~Bièvre}, \citenamefont {Horoshko}, \citenamefont {Patera},\ and\ \citenamefont {Kolobov}}]{De_Bievre2019-se}%
  \BibitemOpen
  \bibfield  {author} {\bibinfo {author} {\bibfnamefont {S.}~\bibnamefont {De~Bièvre}}, \bibinfo {author} {\bibfnamefont {D.~B.}\ \bibnamefont {Horoshko}}, \bibinfo {author} {\bibfnamefont {G.}~\bibnamefont {Patera}},\ and\ \bibinfo {author} {\bibfnamefont {M.~I.}\ \bibnamefont {Kolobov}},\ }\bibfield  {title} {\bibinfo {title} {Measuring nonclassicality of bosonic field quantum states via operator ordering sensitivity},\ }\href {https://doi.org/10.1103/PhysRevLett.122.080402} {\bibfield  {journal} {\bibinfo  {journal} {Phys. Rev. Lett.}\ }\textbf {\bibinfo {volume} {122}},\ \bibinfo {pages} {080402} (\bibinfo {year} {2019})}\BibitemShut {NoStop}%
\bibitem [{\citenamefont {Wigner}(1932)}]{Wigner1932-en}%
  \BibitemOpen
  \bibfield  {author} {\bibinfo {author} {\bibfnamefont {E.}~\bibnamefont {Wigner}},\ }\bibfield  {title} {\bibinfo {title} {On the quantum correction for thermodynamic equilibrium},\ }\href {https://doi.org/10.1103/PhysRev.40.749} {\bibfield  {journal} {\bibinfo  {journal} {Phys. Rev.}\ }\textbf {\bibinfo {volume} {40}},\ \bibinfo {pages} {749} (\bibinfo {year} {1932})}\BibitemShut {NoStop}%
\bibitem [{\citenamefont {Kenfack}\ and\ \citenamefont {Życzkowski}(2004)}]{Kenfack2004-cr}%
  \BibitemOpen
  \bibfield  {author} {\bibinfo {author} {\bibfnamefont {A.}~\bibnamefont {Kenfack}}\ and\ \bibinfo {author} {\bibfnamefont {K.}~\bibnamefont {Życzkowski}},\ }\bibfield  {title} {\bibinfo {title} {Negativity of the {W}igner function as an indicator of non-classicality},\ }\href {https://doi.org/10.1088/1464-4266/6/10/003} {\bibfield  {journal} {\bibinfo  {journal} {J. Opt. B Quantum Semiclassical Opt.}\ }\textbf {\bibinfo {volume} {6}},\ \bibinfo {pages} {396} (\bibinfo {year} {2004})}\BibitemShut {NoStop}%
\bibitem [{\citenamefont {Mari}\ and\ \citenamefont {Eisert}(2012)}]{Mari2012-fs}%
  \BibitemOpen
  \bibfield  {author} {\bibinfo {author} {\bibfnamefont {A.}~\bibnamefont {Mari}}\ and\ \bibinfo {author} {\bibfnamefont {J.}~\bibnamefont {Eisert}},\ }\bibfield  {title} {\bibinfo {title} {Positive {W}igner functions render classical simulation of quantum computation efficient},\ }\href {https://doi.org/10.1103/PhysRevLett.109.230503} {\bibfield  {journal} {\bibinfo  {journal} {Phys. Rev. Lett.}\ }\textbf {\bibinfo {volume} {109}},\ \bibinfo {pages} {230503} (\bibinfo {year} {2012})}\BibitemShut {NoStop}%
\bibitem [{Note1()}]{Note1}%
  \BibitemOpen
  \bibinfo {note} {Strictly speaking, states whose Wigner function does not admit any negativity should be denoted as Wigner-non-negative states (since their Wigner function is $\ge 0$ and may even have zeros), but we prefer denoting them as Wigner-positive states in this paper for brevity.}\BibitemShut {Stop}%
\bibitem [{\citenamefont {Kastler}(1965)}]{Kastler1965-bv}%
  \BibitemOpen
  \bibfield  {author} {\bibinfo {author} {\bibfnamefont {D.}~\bibnamefont {Kastler}},\ }\bibfield  {title} {\bibinfo {title} {The {C}$^{\ast}$-algebras of a free boson field: {I}. discussion of the basic facts},\ }\href {https://doi.org/10.1007/bf01649588} {\bibfield  {journal} {\bibinfo  {journal} {Commun. Math. Phys.}\ }\textbf {\bibinfo {volume} {1}},\ \bibinfo {pages} {14} (\bibinfo {year} {1965})}\BibitemShut {NoStop}%
\bibitem [{\citenamefont {Loupias}\ and\ \citenamefont {Miracle-Sole}(1966)}]{Loupias1966-vd}%
  \BibitemOpen
  \bibfield  {author} {\bibinfo {author} {\bibfnamefont {G.}~\bibnamefont {Loupias}}\ and\ \bibinfo {author} {\bibfnamefont {S.}~\bibnamefont {Miracle-Sole}},\ }\bibfield  {title} {\bibinfo {title} {C$^{\ast}$-algèbres des systèmes canoniques. {I}},\ }\href {https://doi.org/10.1007/bf01773339} {\bibfield  {journal} {\bibinfo  {journal} {Commun. Math. Phys.}\ }\textbf {\bibinfo {volume} {2}},\ \bibinfo {pages} {31} (\bibinfo {year} {1966})}\BibitemShut {NoStop}%
\bibitem [{\citenamefont {Loupias}\ and\ \citenamefont {Miracle-Sole}(1967)}]{Loupias1967-bg}%
  \BibitemOpen
  \bibfield  {author} {\bibinfo {author} {\bibfnamefont {G.}~\bibnamefont {Loupias}}\ and\ \bibinfo {author} {\bibfnamefont {S.}~\bibnamefont {Miracle-Sole}},\ }\bibfield  {title} {\bibinfo {title} {C$^{\ast}$-algèbres des systèmes canoniques. {II}},\ }\href@noop {} {\bibfield  {journal} {\bibinfo  {journal} {Annales de l'institut Henri Poincaré. Section A, Physique Théorique}\ }\textbf {\bibinfo {volume} {6}},\ \bibinfo {pages} {39} (\bibinfo {year} {1967})}\BibitemShut {NoStop}%
\bibitem [{\citenamefont {Narcowich}\ and\ \citenamefont {O'Connell}(1986)}]{Narcowich1986-kq}%
  \BibitemOpen
  \bibfield  {author} {\bibinfo {author} {\bibfnamefont {F.~J.}\ \bibnamefont {Narcowich}}\ and\ \bibinfo {author} {\bibfnamefont {R.~F.}\ \bibnamefont {O'Connell}},\ }\bibfield  {title} {\bibinfo {title} {Necessary and sufficient conditions for a phase-space function to be a {W}igner distribution},\ }\href {https://doi.org/10.1103/physreva.34.1} {\bibfield  {journal} {\bibinfo  {journal} {Phys. Rev. A}\ }\textbf {\bibinfo {volume} {34}},\ \bibinfo {pages} {1} (\bibinfo {year} {1986})}\BibitemShut {NoStop}%
\bibitem [{\citenamefont {Robertson}(1929)}]{Robertson1929-js}%
  \BibitemOpen
  \bibfield  {author} {\bibinfo {author} {\bibfnamefont {H.~P.}\ \bibnamefont {Robertson}},\ }\bibfield  {title} {\bibinfo {title} {The uncertainty principle},\ }\href {https://doi.org/10.1103/physrev.34.163} {\bibfield  {journal} {\bibinfo  {journal} {Phys. Rev.}\ }\textbf {\bibinfo {volume} {34}},\ \bibinfo {pages} {163} (\bibinfo {year} {1929})}\BibitemShut {NoStop}%
\bibitem [{\citenamefont {Białynicki-Birula}\ and\ \citenamefont {Mycielski}(1975)}]{Bialynicki-Birula1975-kv}%
  \BibitemOpen
  \bibfield  {author} {\bibinfo {author} {\bibfnamefont {I.}~\bibnamefont {Białynicki-Birula}}\ and\ \bibinfo {author} {\bibfnamefont {J.}~\bibnamefont {Mycielski}},\ }\bibfield  {title} {\bibinfo {title} {Uncertainty relations for information entropy in wave mechanics},\ }\href {https://doi.org/10.1007/BF01608825} {\bibfield  {journal} {\bibinfo  {journal} {Commun. Math. Phys.}\ }\textbf {\bibinfo {volume} {44}},\ \bibinfo {pages} {129} (\bibinfo {year} {1975})}\BibitemShut {NoStop}%
\bibitem [{\citenamefont {Bialynicki-Birula}(2007)}]{Bialynicki-Birula2007-lp}%
  \BibitemOpen
  \bibfield  {author} {\bibinfo {author} {\bibfnamefont {I.}~\bibnamefont {Bialynicki-Birula}},\ }\bibfield  {title} {\bibinfo {title} {Rényi entropy and the uncertainty relations},\ }\href {https://doi.org/10.1063/1.2713446} {\bibfield  {journal} {\bibinfo  {journal} {Foundations of probability and physics}\ }\textbf {\bibinfo {volume} {889}},\ \bibinfo {pages} {52} (\bibinfo {year} {2007})}\BibitemShut {NoStop}%
\bibitem [{\citenamefont {Van~Herstraeten}\ and\ \citenamefont {Cerf}(2021)}]{Van_Herstraeten2021-nj}%
  \BibitemOpen
  \bibfield  {author} {\bibinfo {author} {\bibfnamefont {Z.}~\bibnamefont {Van~Herstraeten}}\ and\ \bibinfo {author} {\bibfnamefont {N.~J.}\ \bibnamefont {Cerf}},\ }\bibfield  {title} {\bibinfo {title} {Quantum {W}igner entropy},\ }\href {https://doi.org/10.1103/PhysRevA.104.042211} {\bibfield  {journal} {\bibinfo  {journal} {Phys. Rev. A}\ }\textbf {\bibinfo {volume} {104}},\ \bibinfo {pages} {042211} (\bibinfo {year} {2021})}\BibitemShut {NoStop}%
\bibitem [{\citenamefont {Hertz}\ \emph {et~al.}(2017)\citenamefont {Hertz}, \citenamefont {Jabbour},\ and\ \citenamefont {Cerf}}]{Hertz2017-ta}%
  \BibitemOpen
  \bibfield  {author} {\bibinfo {author} {\bibfnamefont {A.}~\bibnamefont {Hertz}}, \bibinfo {author} {\bibfnamefont {M.~G.}\ \bibnamefont {Jabbour}},\ and\ \bibinfo {author} {\bibfnamefont {N.~J.}\ \bibnamefont {Cerf}},\ }\bibfield  {title} {\bibinfo {title} {Entropy-power uncertainty relations: towards a tight inequality for all {G}aussian pure states},\ }\href {https://doi.org/10.1088/1751-8121/aa852f} {\bibfield  {journal} {\bibinfo  {journal} {J. Phys. A: Math. Theor.}\ }\textbf {\bibinfo {volume} {50}},\ \bibinfo {pages} {385301} (\bibinfo {year} {2017})}\BibitemShut {NoStop}%
\bibitem [{\citenamefont {Van~Herstraeten}\ \emph {et~al.}(2023)\citenamefont {Van~Herstraeten}, \citenamefont {Jabbour},\ and\ \citenamefont {Cerf}}]{Van-Herstraeten2023-ww}%
  \BibitemOpen
  \bibfield  {author} {\bibinfo {author} {\bibfnamefont {Z.}~\bibnamefont {Van~Herstraeten}}, \bibinfo {author} {\bibfnamefont {M.~G.}\ \bibnamefont {Jabbour}},\ and\ \bibinfo {author} {\bibfnamefont {N.~J.}\ \bibnamefont {Cerf}},\ }\bibfield  {title} {\bibinfo {title} {Continuous majorization in quantum phase space},\ }\href {https://doi.org/10.22331/q-2023-05-24-1021} {\bibfield  {journal} {\bibinfo  {journal} {Quantum}\ }\textbf {\bibinfo {volume} {7}},\ \bibinfo {pages} {1021} (\bibinfo {year} {2023})}\BibitemShut {NoStop}%
\bibitem [{\citenamefont {Cerf}\ \emph {et~al.}(2023)\citenamefont {Cerf}, \citenamefont {Hertz},\ and\ \citenamefont {Van~Herstraeten}}]{Cerf2023-gn}%
  \BibitemOpen
  \bibfield  {author} {\bibinfo {author} {\bibfnamefont {N.}~\bibnamefont {Cerf}}, \bibinfo {author} {\bibfnamefont {A.}~\bibnamefont {Hertz}},\ and\ \bibinfo {author} {\bibfnamefont {Z.}~\bibnamefont {Van~Herstraeten}},\ }\bibfield  {title} {\bibinfo {title} {Complex-valued {W}igner entropy of a quantum state},\ }\bibfield  {journal} {\bibinfo  {journal} {Quantum Stud. Math. Found.}\ }\href {https://doi.org/10.1007/s40509-024-00325-8} {10.1007/s40509-024-00325-8} (\bibinfo {year} {2023})\BibitemShut {NoStop}%
\bibitem [{\citenamefont {Dias}\ and\ \citenamefont {Prata}(2023)}]{Dias2023-qr}%
  \BibitemOpen
  \bibfield  {author} {\bibinfo {author} {\bibfnamefont {N.~C.}\ \bibnamefont {Dias}}\ and\ \bibinfo {author} {\bibfnamefont {J.~N.}\ \bibnamefont {Prata}},\ }\bibfield  {title} {\bibinfo {title} {On a recent conjecture by {Z}. {V}an {H}erstraeten and {N}. {J}. {C}erf for the quantum {W}igner entropy},\ }\bibfield  {journal} {\bibinfo  {journal} {Annales Henri Poincaré}\ }\href {https://doi.org/10.1007/s00023-023-01298-x} {10.1007/s00023-023-01298-x} (\bibinfo {year} {2023})\BibitemShut {NoStop}%
\bibitem [{\citenamefont {Qian}\ and\ \citenamefont {Gagatsos}(2024)}]{Qian2024-vl}%
  \BibitemOpen
  \bibfield  {author} {\bibinfo {author} {\bibfnamefont {Q.}~\bibnamefont {Qian}}\ and\ \bibinfo {author} {\bibfnamefont {C.~N.}\ \bibnamefont {Gagatsos}},\ }\bibfield  {title} {\bibinfo {title} {Wigner non-negative states that verify the {W}igner entropy conjecture},\ }\href {https://doi.org/10.1103/physreva.110.012228} {\bibfield  {journal} {\bibinfo  {journal} {Phys. Rev. A}\ }\textbf {\bibinfo {volume} {110}},\ \bibinfo {pages} {012228} (\bibinfo {year} {2024})}\BibitemShut {NoStop}%
\bibitem [{\citenamefont {Janssen}(1979)}]{Janssen1979-es}%
  \BibitemOpen
  \bibfield  {author} {\bibinfo {author} {\bibfnamefont {A.~J. E.~M.}\ \bibnamefont {Janssen}},\ }\emph {\bibinfo {title} {Application of the Wigner distribution to harmonic analysis of generalized stochastic processes}},\ \href {https://doi.org/10.6100/IR7013} {Ph.D. thesis},\ \bibinfo  {school} {Technische Hogeschool Eindhoven} (\bibinfo {year} {1979})\BibitemShut {NoStop}%
\bibitem [{\citenamefont {Hlawatsch}(1984)}]{Hlawatsch1984-yx}%
  \BibitemOpen
  \bibfield  {author} {\bibinfo {author} {\bibfnamefont {F.}~\bibnamefont {Hlawatsch}},\ }\bibfield  {title} {\bibinfo {title} {Interference terms in the {W}igner distribution},\ }\href@noop {} {\bibfield  {journal} {\bibinfo  {journal} {Proc. Dig. Sig. Proc.}\ }\textbf {\bibinfo {volume} {363}} (\bibinfo {year} {1984})}\BibitemShut {NoStop}%
\bibitem [{\citenamefont {Lieb}(1990)}]{Lieb1990-ev}%
  \BibitemOpen
  \bibfield  {author} {\bibinfo {author} {\bibfnamefont {E.~H.}\ \bibnamefont {Lieb}},\ }\bibfield  {title} {\bibinfo {title} {Integral bounds for radar ambiguity functions and {W}igner distributions},\ }\href {https://doi.org/10.1063/1.528894} {\bibfield  {journal} {\bibinfo  {journal} {J. Math. Phys.}\ }\textbf {\bibinfo {volume} {31}},\ \bibinfo {pages} {594} (\bibinfo {year} {1990})}\BibitemShut {NoStop}%
\bibitem [{\citenamefont {Royer}(1977)}]{Royer1977-he}%
  \BibitemOpen
  \bibfield  {author} {\bibinfo {author} {\bibfnamefont {A.}~\bibnamefont {Royer}},\ }\bibfield  {title} {\bibinfo {title} {Wigner function as the expectation value of a parity operator},\ }\href {https://doi.org/10.1103/PhysRevA.15.449} {\bibfield  {journal} {\bibinfo  {journal} {Phys. Rev. A}\ }\textbf {\bibinfo {volume} {15}},\ \bibinfo {pages} {449} (\bibinfo {year} {1977})}\BibitemShut {NoStop}%
\bibitem [{Note2()}]{Note2}%
  \BibitemOpen
  \bibinfo {note} {Note the factor $\protect \sqrt {2}$ in Eqs. \protect \eqref {eq:weyl_transform} and \protect \eqref {eq:inverse_weyl_transform}, which originates from our asymmetric conventions, namely $\protect \hat {a}=(\protect \hat {x}+i\protect \hat {p})/\protect \sqrt {2}$ but $\alpha =x+ip$. It implies that the displacement operator $\protect \hat {D}(\alpha )$ shifts the $(x,p)$ coordinates as $x\mapsto x+ \protect \sqrt {2} \protect \,\protect \mathrm {Re(\alpha )}$ and $p\mapsto p + \protect \sqrt {2} \protect \,\protect \mathrm {Im(\alpha )}$. The reason for this choice is that the definition \protect \eqref {eq-def-Wigner-entropy-phase-space} of the Wigner entropy then coincides with its earlier definition in terms of $(x,p)$ coordinates, see \cite {Van_Herstraeten2021-nj}.}\BibitemShut {Stop}%
\bibitem [{\citenamefont {Moyal}(1949)}]{Moyal1949-tl}%
  \BibitemOpen
  \bibfield  {author} {\bibinfo {author} {\bibfnamefont {J.~E.}\ \bibnamefont {Moyal}},\ }\bibfield  {title} {\bibinfo {title} {Quantum mechanics as a statistical theory},\ }\href {https://doi.org/10.1017/S0305004100000487} {\bibfield  {journal} {\bibinfo  {journal} {Math. Proc. Cambridge Philos. Soc.}\ }\textbf {\bibinfo {volume} {45}},\ \bibinfo {pages} {99} (\bibinfo {year} {1949})}\BibitemShut {NoStop}%
\bibitem [{\citenamefont {Hudson}(1974)}]{Hudson1974-ll}%
  \BibitemOpen
  \bibfield  {author} {\bibinfo {author} {\bibfnamefont {R.~L.}\ \bibnamefont {Hudson}},\ }\bibfield  {title} {\bibinfo {title} {When is the {W}igner quasi-probability density non-negative?},\ }\href {https://doi.org/10.1016/0034-4877(74)90007-X} {\bibfield  {journal} {\bibinfo  {journal} {Rep. Math. Phys.}\ }\textbf {\bibinfo {volume} {6}},\ \bibinfo {pages} {249} (\bibinfo {year} {1974})}\BibitemShut {NoStop}%
\bibitem [{\citenamefont {Garcia-Bondia}\ and\ \citenamefont {Várilly}(1988)}]{Garcia-Bondia1988-tr}%
  \BibitemOpen
  \bibfield  {author} {\bibinfo {author} {\bibfnamefont {J.}~\bibnamefont {Garcia-Bondia}}\ and\ \bibinfo {author} {\bibfnamefont {J.~C.}\ \bibnamefont {Várilly}},\ }\bibfield  {title} {\bibinfo {title} {Nonnegative mixed states in {W}eyl-{W}igner-{M}oyal theory},\ }\href {https://doi.org/10.1016/0375-9601(88)91035-3} {\bibfield  {journal} {\bibinfo  {journal} {Phys. Lett. A}\ }\textbf {\bibinfo {volume} {128}},\ \bibinfo {pages} {20} (\bibinfo {year} {1988})}\BibitemShut {NoStop}%
\bibitem [{\citenamefont {Bröcker}\ and\ \citenamefont {Werner}(1995)}]{Brocker1995-rp}%
  \BibitemOpen
  \bibfield  {author} {\bibinfo {author} {\bibfnamefont {T.}~\bibnamefont {Bröcker}}\ and\ \bibinfo {author} {\bibfnamefont {R.~F.}\ \bibnamefont {Werner}},\ }\bibfield  {title} {\bibinfo {title} {Mixed states with positive {W}igner functions},\ }\href {https://doi.org/10.1063/1.531326} {\bibfield  {journal} {\bibinfo  {journal} {J. Math. Phys.}\ }\textbf {\bibinfo {volume} {36}},\ \bibinfo {pages} {62} (\bibinfo {year} {1995})}\BibitemShut {NoStop}%
\bibitem [{\citenamefont {Mandilara}\ \emph {et~al.}(2009)\citenamefont {Mandilara}, \citenamefont {Karpov},\ and\ \citenamefont {Cerf}}]{Mandilara2009-kn}%
  \BibitemOpen
  \bibfield  {author} {\bibinfo {author} {\bibfnamefont {A.}~\bibnamefont {Mandilara}}, \bibinfo {author} {\bibfnamefont {E.}~\bibnamefont {Karpov}},\ and\ \bibinfo {author} {\bibfnamefont {N.~J.}\ \bibnamefont {Cerf}},\ }\bibfield  {title} {\bibinfo {title} {Extending {H}udson's theorem to mixed quantum states},\ }\href {https://doi.org/10.1103/PhysRevA.79.062302} {\bibfield  {journal} {\bibinfo  {journal} {Phys. Rev. A}\ }\textbf {\bibinfo {volume} {79}},\ \bibinfo {pages} {062302} (\bibinfo {year} {2009})}\BibitemShut {NoStop}%
\bibitem [{\citenamefont {Lenard}(1978)}]{Lenard1978-ru}%
  \BibitemOpen
  \bibfield  {author} {\bibinfo {author} {\bibfnamefont {A.}~\bibnamefont {Lenard}},\ }\bibfield  {title} {\bibinfo {title} {Thermodynamical proof of the {G}ibbs formula for elementary quantum systems},\ }\href {https://doi.org/10.1007/BF01011769} {\bibfield  {journal} {\bibinfo  {journal} {J. Stat. Phys.}\ }\textbf {\bibinfo {volume} {19}},\ \bibinfo {pages} {575} (\bibinfo {year} {1978})}\BibitemShut {NoStop}%
\bibitem [{\citenamefont {Bastiaans}(1983)}]{Bastiaans1983-je}%
  \BibitemOpen
  \bibfield  {author} {\bibinfo {author} {\bibfnamefont {M.~J.}\ \bibnamefont {Bastiaans}},\ }\bibfield  {title} {\bibinfo {title} {Lower bound in the uncertainty principle for partially coherent light},\ }\href {https://doi.org/10.1364/JOSA.73.001320} {\bibfield  {journal} {\bibinfo  {journal} {J. Opt. Soc. Am., JOSA}\ }\textbf {\bibinfo {volume} {73}},\ \bibinfo {pages} {1320} (\bibinfo {year} {1983})}\BibitemShut {NoStop}%
\bibitem [{\citenamefont {Santos}\ \emph {et~al.}(2017)\citenamefont {Santos}, \citenamefont {Landi},\ and\ \citenamefont {Paternostro}}]{PhysRevLett.118.220601}%
  \BibitemOpen
  \bibfield  {author} {\bibinfo {author} {\bibfnamefont {J.~P.}\ \bibnamefont {Santos}}, \bibinfo {author} {\bibfnamefont {G.~T.}\ \bibnamefont {Landi}},\ and\ \bibinfo {author} {\bibfnamefont {M.}~\bibnamefont {Paternostro}},\ }\bibfield  {title} {\bibinfo {title} {Wigner entropy production rate},\ }\href {https://doi.org/10.1103/PhysRevLett.118.220601} {\bibfield  {journal} {\bibinfo  {journal} {Phys. Rev. Lett.}\ }\textbf {\bibinfo {volume} {118}},\ \bibinfo {pages} {220601} (\bibinfo {year} {2017})}\BibitemShut {NoStop}%
\bibitem [{\citenamefont {Brunelli}\ \emph {et~al.}(2018)\citenamefont {Brunelli}, \citenamefont {Fusco}, \citenamefont {Landig}, \citenamefont {Wieczorek}, \citenamefont {Hoelscher-Obermaier}, \citenamefont {Landi}, \citenamefont {Semi\~ao}, \citenamefont {Ferraro}, \citenamefont {Kiesel}, \citenamefont {Donner}, \citenamefont {De~Chiara},\ and\ \citenamefont {Paternostro}}]{PhysRevLett.121.160604}%
  \BibitemOpen
  \bibfield  {author} {\bibinfo {author} {\bibfnamefont {M.}~\bibnamefont {Brunelli}}, \bibinfo {author} {\bibfnamefont {L.}~\bibnamefont {Fusco}}, \bibinfo {author} {\bibfnamefont {R.}~\bibnamefont {Landig}}, \bibinfo {author} {\bibfnamefont {W.}~\bibnamefont {Wieczorek}}, \bibinfo {author} {\bibfnamefont {J.}~\bibnamefont {Hoelscher-Obermaier}}, \bibinfo {author} {\bibfnamefont {G.}~\bibnamefont {Landi}}, \bibinfo {author} {\bibfnamefont {F.~L.}\ \bibnamefont {Semi\~ao}}, \bibinfo {author} {\bibfnamefont {A.}~\bibnamefont {Ferraro}}, \bibinfo {author} {\bibfnamefont {N.}~\bibnamefont {Kiesel}}, \bibinfo {author} {\bibfnamefont {T.}~\bibnamefont {Donner}}, \bibinfo {author} {\bibfnamefont {G.}~\bibnamefont {De~Chiara}},\ and\ \bibinfo {author} {\bibfnamefont {M.}~\bibnamefont {Paternostro}},\ }\bibfield  {title} {\bibinfo {title} {Experimental determination of irreversible entropy production in out-of-equilibrium mesoscopic quantum systems},\ }\href {https://doi.org/10.1103/PhysRevLett.121.160604} {\bibfield  {journal} {\bibinfo  {journal} {Phys. Rev. Lett.}\ }\textbf {\bibinfo {volume} {121}},\ \bibinfo {pages} {160604} (\bibinfo {year} {2018})}\BibitemShut {NoStop}%
\bibitem [{\citenamefont {Adesso}\ \emph {et~al.}(2012)\citenamefont {Adesso}, \citenamefont {Girolami},\ and\ \citenamefont {Serafini}}]{Adesso2012-ok}%
  \BibitemOpen
  \bibfield  {author} {\bibinfo {author} {\bibfnamefont {G.}~\bibnamefont {Adesso}}, \bibinfo {author} {\bibfnamefont {D.}~\bibnamefont {Girolami}},\ and\ \bibinfo {author} {\bibfnamefont {A.}~\bibnamefont {Serafini}},\ }\bibfield  {title} {\bibinfo {title} {Measuring gaussian quantum information and correlations using the rényi entropy of order 2},\ }\href {https://doi.org/10.1103/PhysRevLett.109.190502} {\bibfield  {journal} {\bibinfo  {journal} {Phys. Rev. Lett.}\ }\textbf {\bibinfo {volume} {109}},\ \bibinfo {pages} {190502} (\bibinfo {year} {2012})}\BibitemShut {NoStop}%
\bibitem [{\citenamefont {Guevara}\ \emph {et~al.}(2003)\citenamefont {Guevara}, \citenamefont {Sagar},\ and\ \citenamefont {Esquivel}}]{Guevara2003-na}%
  \BibitemOpen
  \bibfield  {author} {\bibinfo {author} {\bibfnamefont {N.~L.}\ \bibnamefont {Guevara}}, \bibinfo {author} {\bibfnamefont {R.~P.}\ \bibnamefont {Sagar}},\ and\ \bibinfo {author} {\bibfnamefont {R.~O.}\ \bibnamefont {Esquivel}},\ }\bibfield  {title} {\bibinfo {title} {Information uncertainty-type inequalities in atomic systems},\ }\href {https://doi.org/10.1063/1.1605932} {\bibfield  {journal} {\bibinfo  {journal} {J. Chem. Phys.}\ }\textbf {\bibinfo {volume} {119}},\ \bibinfo {pages} {7030} (\bibinfo {year} {2003})}\BibitemShut {NoStop}%
\bibitem [{\citenamefont {Laguna}\ and\ \citenamefont {Sagar}(2010)}]{Laguna2010-ir}%
  \BibitemOpen
  \bibfield  {author} {\bibinfo {author} {\bibfnamefont {H.~G.}\ \bibnamefont {Laguna}}\ and\ \bibinfo {author} {\bibfnamefont {R.~P.}\ \bibnamefont {Sagar}},\ }\bibfield  {title} {\bibinfo {title} {Shannon entropy of the {W}igner function and position-momentum correlation in model systems},\ }\href {https://doi.org/10.1142/S0219749910006484} {\bibfield  {journal} {\bibinfo  {journal} {Int. J. Quantum Inform.}\ }\textbf {\bibinfo {volume} {08}},\ \bibinfo {pages} {1089} (\bibinfo {year} {2010})}\BibitemShut {NoStop}%
\bibitem [{\citenamefont {Salazar}\ \emph {et~al.}(2023)\citenamefont {Salazar}, \citenamefont {Laguna},\ and\ \citenamefont {Sagar}}]{Salazar2023-yf}%
  \BibitemOpen
  \bibfield  {author} {\bibinfo {author} {\bibfnamefont {S.~J.~C.}\ \bibnamefont {Salazar}}, \bibinfo {author} {\bibfnamefont {H.~G.}\ \bibnamefont {Laguna}},\ and\ \bibinfo {author} {\bibfnamefont {R.~P.}\ \bibnamefont {Sagar}},\ }\bibfield  {title} {\bibinfo {title} {Phase-space quantum distributions and information theory},\ }\href {https://doi.org/10.1103/PhysRevA.107.042417} {\bibfield  {journal} {\bibinfo  {journal} {Phys. Rev. A}\ }\textbf {\bibinfo {volume} {107}},\ \bibinfo {pages} {042417} (\bibinfo {year} {2023})}\BibitemShut {NoStop}%
\bibitem [{Note3()}]{Note3}%
  \BibitemOpen
  \bibinfo {note} {The lower bound $\ln (2\pi )+ S(\protect \hat \rho )$ depends on the von Neumann entropy $S(\protect \hat \rho )$ of state $\protect \hat \rho $. [T. Haas, private communication, 2024].}\BibitemShut {Stop}%
\bibitem [{\citenamefont {Janssen}(1998)}]{Janssen1998-as}%
  \BibitemOpen
  \bibfield  {author} {\bibinfo {author} {\bibfnamefont {A.~J. E.~M.}\ \bibnamefont {Janssen}},\ }\bibfield  {title} {\bibinfo {title} {Proof of a conjecture on the supports of {W}igner distributions},\ }\href {https://doi.org/10.1007/BF02479675} {\bibfield  {journal} {\bibinfo  {journal} {J. Fourier Anal. Appl.}\ }\textbf {\bibinfo {volume} {4}},\ \bibinfo {pages} {723} (\bibinfo {year} {1998})}\BibitemShut {NoStop}%
\bibitem [{\citenamefont {Weedbrook}\ \emph {et~al.}(2012)\citenamefont {Weedbrook}, \citenamefont {Pirandola}, \citenamefont {García-Patrón}, \citenamefont {Cerf}, \citenamefont {Ralph}, \citenamefont {Shapiro},\ and\ \citenamefont {Lloyd}}]{Weedbrook2012-qu}%
  \BibitemOpen
  \bibfield  {author} {\bibinfo {author} {\bibfnamefont {C.}~\bibnamefont {Weedbrook}}, \bibinfo {author} {\bibfnamefont {S.}~\bibnamefont {Pirandola}}, \bibinfo {author} {\bibfnamefont {R.}~\bibnamefont {García-Patrón}}, \bibinfo {author} {\bibfnamefont {N.~J.}\ \bibnamefont {Cerf}}, \bibinfo {author} {\bibfnamefont {T.~C.}\ \bibnamefont {Ralph}}, \bibinfo {author} {\bibfnamefont {J.~H.}\ \bibnamefont {Shapiro}},\ and\ \bibinfo {author} {\bibfnamefont {S.}~\bibnamefont {Lloyd}},\ }\bibfield  {title} {\bibinfo {title} {Gaussian quantum information},\ }\href {https://doi.org/10.1103/RevModPhys.84.621} {\bibfield  {journal} {\bibinfo  {journal} {Rev. Mod. Phys.}\ }\textbf {\bibinfo {volume} {84}},\ \bibinfo {pages} {621} (\bibinfo {year} {2012})}\BibitemShut {NoStop}%
\bibitem [{\citenamefont {Bertrand}\ \emph {et~al.}(1983)\citenamefont {Bertrand}, \citenamefont {Doremus}, \citenamefont {Izrar}, \citenamefont {Nguyen},\ and\ \citenamefont {Feix}}]{Bertrand1983-ij}%
  \BibitemOpen
  \bibfield  {author} {\bibinfo {author} {\bibfnamefont {P.}~\bibnamefont {Bertrand}}, \bibinfo {author} {\bibfnamefont {J.~P.}\ \bibnamefont {Doremus}}, \bibinfo {author} {\bibfnamefont {B.}~\bibnamefont {Izrar}}, \bibinfo {author} {\bibfnamefont {V.~T.}\ \bibnamefont {Nguyen}},\ and\ \bibinfo {author} {\bibfnamefont {M.~R.}\ \bibnamefont {Feix}},\ }\bibfield  {title} {\bibinfo {title} {Obtaining non-negative quantum mechanical distribution function},\ }\href {https://doi.org/10.1016/0375-9601(83)90841-1} {\bibfield  {journal} {\bibinfo  {journal} {Phys. Lett. A}\ }\textbf {\bibinfo {volume} {94}},\ \bibinfo {pages} {415} (\bibinfo {year} {1983})}\BibitemShut {NoStop}%
\bibitem [{\citenamefont {Jagannathan}\ \emph {et~al.}(1987)\citenamefont {Jagannathan}, \citenamefont {Simon}, \citenamefont {Sudarshan},\ and\ \citenamefont {Vasudevan}}]{Jagannathan1987-yj}%
  \BibitemOpen
  \bibfield  {author} {\bibinfo {author} {\bibfnamefont {R.}~\bibnamefont {Jagannathan}}, \bibinfo {author} {\bibfnamefont {R.}~\bibnamefont {Simon}}, \bibinfo {author} {\bibfnamefont {E.~C.~G.}\ \bibnamefont {Sudarshan}},\ and\ \bibinfo {author} {\bibfnamefont {R.}~\bibnamefont {Vasudevan}},\ }\bibfield  {title} {\bibinfo {title} {Dynamical maps and nonnegative phase-space distribution functions in quantum mechanics},\ }\href {https://doi.org/10.1016/0375-9601(87)90327-6} {\bibfield  {journal} {\bibinfo  {journal} {Phys. Lett. A}\ }\textbf {\bibinfo {volume} {120}},\ \bibinfo {pages} {161} (\bibinfo {year} {1987})}\BibitemShut {NoStop}%
\bibitem [{\citenamefont {Narcowich}(1988)}]{Narcowich1988-jr}%
  \BibitemOpen
  \bibfield  {author} {\bibinfo {author} {\bibfnamefont {F.~J.}\ \bibnamefont {Narcowich}},\ }\bibfield  {title} {\bibinfo {title} {Conditions for the convolution of two {W}igner distributions to be itself a {W}igner distribution},\ }\href {https://doi.org/10.1063/1.527861} {\bibfield  {journal} {\bibinfo  {journal} {J. Math. Phys.}\ }\textbf {\bibinfo {volume} {29}},\ \bibinfo {pages} {2036} (\bibinfo {year} {1988})}\BibitemShut {NoStop}%
\bibitem [{Note4()}]{Note4}%
  \BibitemOpen
  \bibinfo {note} {Note the discrepancy with the notation used in Ref.~\cite {Van_Herstraeten2021-nj}, where $\protect \mathcal {B}$ denotes the subset of states $\protect \hat {\sigma }(\protect \hat {\rho }_1,\protect \hat {\rho }_2)$ whereas $\protect \mathcal {B}_c$ denotes the closure of its convex hull.}\BibitemShut {Stop}%
\bibitem [{\citenamefont {Lieb}(1978)}]{Lieb1978-gy}%
  \BibitemOpen
  \bibfield  {author} {\bibinfo {author} {\bibfnamefont {E.~H.}\ \bibnamefont {Lieb}},\ }\bibfield  {title} {\bibinfo {title} {Proof of an entropy conjecture of {W}ehrl},\ }\href {https://doi.org/10.1007/BF01940328} {\bibfield  {journal} {\bibinfo  {journal} {Commun. Math. Phys.}\ }\textbf {\bibinfo {volume} {62}},\ \bibinfo {pages} {35} (\bibinfo {year} {1978})}\BibitemShut {NoStop}%
\bibitem [{\citenamefont {Lieb}\ and\ \citenamefont {Solovej}(2014)}]{Lieb2014-qx}%
  \BibitemOpen
  \bibfield  {author} {\bibinfo {author} {\bibfnamefont {E.~H.}\ \bibnamefont {Lieb}}\ and\ \bibinfo {author} {\bibfnamefont {J.~P.}\ \bibnamefont {Solovej}},\ }\bibfield  {title} {\bibinfo {title} {Proof of an entropy conjecture for {B}loch coherent spin states and its generalizations},\ }\href {https://doi.org/10.1007/s11511-014-0113-6} {\bibfield  {journal} {\bibinfo  {journal} {Acta Math.}\ }\textbf {\bibinfo {volume} {212}},\ \bibinfo {pages} {379} (\bibinfo {year} {2014})}\BibitemShut {NoStop}%
\bibitem [{\citenamefont {Grossmann}(1976)}]{Grossmann1976-vk}%
  \BibitemOpen
  \bibfield  {author} {\bibinfo {author} {\bibfnamefont {A.}~\bibnamefont {Grossmann}},\ }\bibfield  {title} {\bibinfo {title} {Parity operator and quantization of $\delta$-functions},\ }\href {https://doi.org/10.1007/BF01617867} {\bibfield  {journal} {\bibinfo  {journal} {Commun. Math. Phys.}\ }\textbf {\bibinfo {volume} {48}},\ \bibinfo {pages} {191} (\bibinfo {year} {1976})}\BibitemShut {NoStop}%
\bibitem [{\citenamefont {Potoček}\ and\ \citenamefont {Barnett}(2015)}]{Potocek2015-ds}%
  \BibitemOpen
  \bibfield  {author} {\bibinfo {author} {\bibfnamefont {V.}~\bibnamefont {Potoček}}\ and\ \bibinfo {author} {\bibfnamefont {S.~M.}\ \bibnamefont {Barnett}},\ }\bibfield  {title} {\bibinfo {title} {On the exponential form of the displacement operator for different systems},\ }\href {https://doi.org/10.1088/0031-8949/90/6/065208} {\bibfield  {journal} {\bibinfo  {journal} {Phys. Scr.}\ }\textbf {\bibinfo {volume} {90}},\ \bibinfo {pages} {065208} (\bibinfo {year} {2015})}\BibitemShut {NoStop}%
\bibitem [{\citenamefont {Janssen}(1982)}]{Janssen1982-op}%
  \BibitemOpen
  \bibfield  {author} {\bibinfo {author} {\bibfnamefont {A.}~\bibnamefont {Janssen}},\ }\bibfield  {title} {\bibinfo {title} {On the locus and spread of pseudo-density functions in the time-frequency plane},\ }\href@noop {} {\bibfield  {journal} {\bibinfo  {journal} {Philips J. Res.}\ }\textbf {\bibinfo {volume} {37}},\ \bibinfo {pages} {79} (\bibinfo {year} {1982})}\BibitemShut {NoStop}%
\bibitem [{\citenamefont {Wehrl}(1979)}]{wehrl1979relation}%
  \BibitemOpen
  \bibfield  {author} {\bibinfo {author} {\bibfnamefont {A.}~\bibnamefont {Wehrl}},\ }\bibfield  {title} {\bibinfo {title} {On the relation between classical and quantum-mechanical entropy},\ }\href@noop {} {\bibfield  {journal} {\bibinfo  {journal} {Reports on Mathematical Physics}\ }\textbf {\bibinfo {volume} {16}},\ \bibinfo {pages} {353} (\bibinfo {year} {1979})}\BibitemShut {NoStop}%
\bibitem [{\citenamefont {Giovannetti}\ \emph {et~al.}(2004)\citenamefont {Giovannetti}, \citenamefont {Guha}, \citenamefont {Lloyd}, \citenamefont {Maccone},\ and\ \citenamefont {Shapiro}}]{PhysRevA.70.032315}%
  \BibitemOpen
  \bibfield  {author} {\bibinfo {author} {\bibfnamefont {V.}~\bibnamefont {Giovannetti}}, \bibinfo {author} {\bibfnamefont {S.}~\bibnamefont {Guha}}, \bibinfo {author} {\bibfnamefont {S.}~\bibnamefont {Lloyd}}, \bibinfo {author} {\bibfnamefont {L.}~\bibnamefont {Maccone}},\ and\ \bibinfo {author} {\bibfnamefont {J.~H.}\ \bibnamefont {Shapiro}},\ }\bibfield  {title} {\bibinfo {title} {Minimum output entropy of bosonic channels: A conjecture},\ }\href {https://doi.org/10.1103/PhysRevA.70.032315} {\bibfield  {journal} {\bibinfo  {journal} {Phys. Rev. A}\ }\textbf {\bibinfo {volume} {70}},\ \bibinfo {pages} {032315} (\bibinfo {year} {2004})}\BibitemShut {NoStop}%
\bibitem [{\citenamefont {Giovannetti}\ \emph {et~al.}(2014)\citenamefont {Giovannetti}, \citenamefont {Garc{\'i}a-Patr{\'o}n}, \citenamefont {Cerf},\ and\ \citenamefont {Holevo}}]{Giovannetti2014}%
  \BibitemOpen
  \bibfield  {author} {\bibinfo {author} {\bibfnamefont {V.}~\bibnamefont {Giovannetti}}, \bibinfo {author} {\bibfnamefont {R.}~\bibnamefont {Garc{\'i}a-Patr{\'o}n}}, \bibinfo {author} {\bibfnamefont {N.~J.}\ \bibnamefont {Cerf}},\ and\ \bibinfo {author} {\bibfnamefont {A.~S.}\ \bibnamefont {Holevo}},\ }\bibfield  {title} {\bibinfo {title} {Ultimate classical communication rates of quantum optical channels},\ }\href {https://doi.org/10.1038/nphoton.2014.216} {\bibfield  {journal} {\bibinfo  {journal} {Nature Photonics}\ }\textbf {\bibinfo {volume} {8}},\ \bibinfo {pages} {796} (\bibinfo {year} {2014})}\BibitemShut {NoStop}%
\end{thebibliography}%

\end{document}